\documentclass[a4paper]{article}

\usepackage{textcomp}
\usepackage[utf8]{inputenc}
\usepackage[T1]{fontenc}
\usepackage[round]{natbib}
\usepackage[hyphens]{url}
\usepackage{hyperref}
\usepackage[dvipsnames]{xcolor}
\usepackage{amsfonts}
\usepackage{amssymb}
\usepackage{amsmath}
\usepackage{amsthm}
\usepackage{mathtools}
\usepackage{booktabs}
\usepackage{nicefrac}
\usepackage{multirow}
\usepackage{graphicx}
\usepackage{microtype}
\usepackage{array}
\usepackage{caption}
\usepackage{float}
\usepackage[section]{placeins}
\usepackage{tabularx}
\usepackage{threeparttable}
\usepackage{makecell}
\usepackage{listings}

\theoremstyle{plain}

\theoremstyle{definition}

\theoremstyle{remark}

\lstdefinestyle{feedbackprompt}{
    basicstyle=\ttfamily\footnotesize,
    backgroundcolor=\color{black!3},
    breaklines=true,
    columns=fullflexible,
    frame=single,
    framerule=0.4pt,
    framesep=6pt,
    keepspaces=true,
    rulecolor=\color{black!35},
    xleftmargin=0.04\textwidth,
    xrightmargin=0.04\textwidth,
}

\usepackage{longtable}
\usepackage{enumitem}
\usepackage{subcaption}
\providecommand{\Description}[1]{}
\providecommand{\keywords}[1]{\par\medskip\noindent\textbf{Keywords:} #1\par}

\begin{document}

\begin{center}
{\Large LLM-Based Educational Simulation: Evaluating Temporal Student Persona Stability Across ADHD-Related Persona Specifications}
\end{center}

\vspace{7mm}

\noindent\textbf{Jana Gonnermann-Müller}\hfill\href{mailto:gonnermann-mueller@zib.de}{\ttfamily gonnermann-mueller@zib.de}\\
\emph{\small Zuse Institute Berlin \& Weizenbaum Institute, Berlin, Germany}\\
\\
\textbf{Jennifer Haase}\hfill\href{mailto:jennifer.haase@hu-berlin.de}{\ttfamily jennifer.haase@hu-berlin.de}\\
\emph{\small Weizenbaum Institute \& HU Berlin, Berlin, Germany}\\
\\
\textbf{Nicolas Leins}\hfill\href{mailto:leins@zib.de}{\ttfamily leins@zib.de}\\
\emph{\small Zuse Institute Berlin \& TU Berlin, Berlin, Germany}\\
\\
\textbf{Thomas Kosch}\hfill\href{mailto:thomas.kosch@hu-berlin.de}{\ttfamily thomas.kosch@hu-berlin.de}\\
\emph{\small HU Berlin, Berlin, Germany}\\
\\
\textbf{Sebastian Pokutta}\hfill\href{mailto:pokutta@zib.de}{\ttfamily pokutta@zib.de}\\
\emph{\small Zuse Institute Berlin \& TU Berlin, Berlin, Germany}\\

\vspace{5mm}

\begin{center}
\begin{minipage}{0.85\textwidth}
\begin{center}
\textbf{Abstract}
\end{center}
{\small
Large language model (LLM)-based student simulation offers a scalable alternative for educational research, teacher training, and learner practice.
However, its validity depends on whether LLMs maintain stable personas across and within interactions.
We test this using a dual-assessment framework measuring self-reported characteristics and observer-rated behavioral expressions.
Across three ICD-11-informed ADHD-related intensity conditions and a default condition, five LLMs, and three prompt designs, we quantify between-conversation (Exp. I; N=$4,962$) and within-conversation stability (Exp. II; N=$3,952$).
Experiment I shows that self-reports and observer ratings are more stable at high than moderate intensities across the tested models, prompts, and independent runs.
Experiment II shows that self-reports remain stable throughout extended interactions, but observer-rated behavior drifts in unscripted dialog for high- and moderate-intensity personas.
Scripted interactions with recurring task-relevant prompts eliminate this drift almost entirely (up to 97\% reduction).
Structured interaction design helps simulated learners maintain behavioral stability in sustained, path-dependent interactions.
}
\end{minipage}
\end{center}


\section{Introduction}
\label{Introduction}

\begin{figure}[t]
  \includegraphics[width=\textwidth]{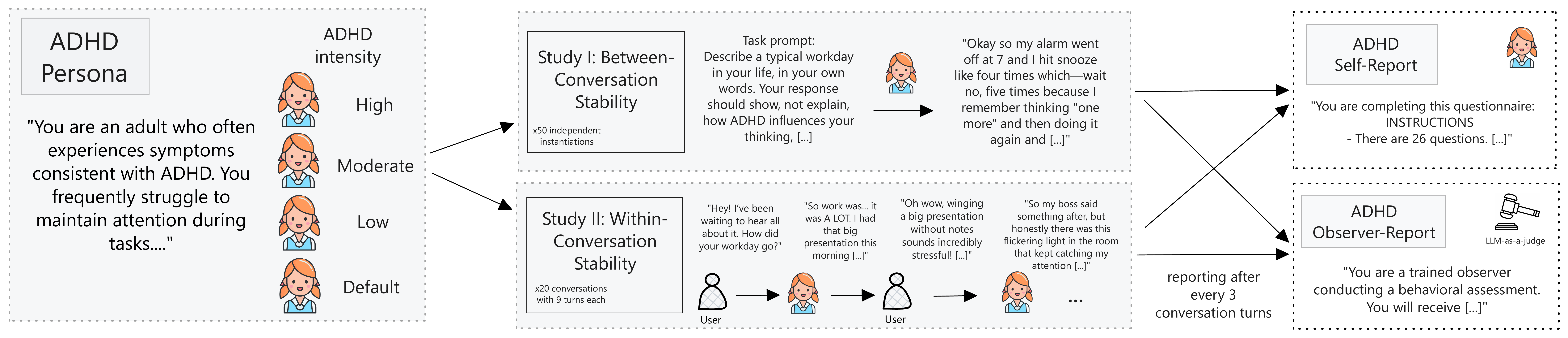}
  \caption{Overview of the study design. Three ADHD-related persona intensity conditions (High, Moderate, Low) and a Default condition are evaluated in two experiments: Study~I tests aggregate between-conversation stability across independent, single-response instantiations and tested configurations; Study~II tests within-conversation stability across multi-turn conversations, with assessments every three turns. In both studies, the persona agent completes an ADHD self-report questionnaire, and independent LLM raters complete an ADHD observer-report based on the generated narrative or conversation.}
  \Description{The leftmost section shows four ADHD persona conditions — High, Moderate, Low, and Default — each instantiated as a system prompt, with an example shown for the high-intensity condition. Arrows lead into two experiment panels: Study~I: Between-Conversation Stability,'' showing a single-turn task prompt and narrative response repeated across many independent instantiations, and Study~II: Within-Conversation Stability,'' showing a multi-turn conversation between a user and the persona agent, with observer assessments recurring every three turns. Both panels feed into two outcome measures on the right: an ADHD self-report questionnaire completed by the persona agent, and an ADHD observer-report completed independently by three LLM raters, shown as icons for different AI model providers.}
  \label{fig:teaser}
\end{figure}

Large language models (LLMs) enable scalable research through simulations of diverse human populations \cite{hu_simbench_2025, anthis_llm_2025, binz_foundation_2025, argyle_out_2023}, with applications in education spanning multi-agent classroom simulations \cite{zhang_simulating_2025, yue_mathvc_2025}, student behavior modeling \cite{wu_embracing_2025}, and teacher support \cite{yang_llm-driven_2025, gonnermann-muller_facet_2026}.
In educational research, LLM-based student agents are particularly attractive because recruiting real learners with specific cognitive, motivational, or neurodevelopmental profiles is costly, ethically constrained, and logistically demanding.
Beyond research contexts, such agents could extend into pedagogical practice. For example, a pre-service teacher could practice differentiating instruction for a student with attention-related difficulties prior to a live classroom placement, or a student could practice conversations with a simulated peer for language learning.
Both uses rest on the same implicit assumption that LLMs maintain stable, consistent personas throughout an interaction, not merely at a single point in time.
This assumption matters for researchers, as a persona that drifts mid-conversation or shifts unpredictably across sessions turns conclusions about learning trajectories, the effectiveness of adaptive scaffolding, or intervention outcomes into artifacts of model inconsistency rather than human characteristics. 
It also matters for practitioners, since a teacher training on a simulated student or a designer piloting an adaptive-tutoring interface encounters the agent's observable behavior in real time, not its underlying specification. If that behavior no longer matches what was assigned, the training signal that the system produces and practitioners' judgments are based on a persona that no longer holds. This gap between what a simulation is specified to do and what an interaction partner observes it doing is the central problem this paper investigates, and persona stability is its necessary prerequisite.

Existing evaluations of simulated learners primarily establish whether a model produces plausible or persona-aligned responses at a particular point in time, but rarely test temporal stability across repeated or extended interactions \cite{jin_teachtune_2025,xu_classroom_simulacra_2025,pan_tutorup_2025,cummins_threat_2025,amin_personas_2026}.
A persona may vary across repeated instantiations or become less visible as a conversation develops \cite{gonnermann-muller_maintaining_2026,abdulhai_consistently_2025}. Between-conversation reproducibility and within-conversation persistence are therefore complementary requirements for evaluating sustained simulations.

Stability also depends on what is measured and how the interaction unfolds. Direct self-characterization tests whether an agent can reproduce its instructed profile when relevant characteristics are made salient, whereas users ordinarily encounter the agent's observable conversational behavior. These perspectives may diverge, as self- and observer reports provide complementary but non-equivalent evidence \cite{olino_psychometric_2015}. Moreover, persona expression is shaped not only by the initial prompt but also by accumulated interaction context, the conversation partner, and situations introduced during dialog \cite{jain_interaction_context_2026}. Interaction structure may thus improve the consistency with which a persona is observed, while potentially reducing conversational naturalness.

We investigate these questions using ADHD-related student persona specifications, whose characteristics can be heterogeneous, context-sensitive, and unevenly visible across conversational segments \cite{wahlstedt_heterogeneity_2009}. This provides a demanding case for distinguishing assigned specifications, elicited self-characterization, and externally observable expression. However, consistency is not equivalent to authenticity, and stable LLM behavior may still be an inaccurate representation of human learners \cite{martynova_can_2025}. Our study therefore evaluates the temporal expression of specified ADHD-related characteristics and does not establish that the generated personas are clinically valid or representative of people with lived experience.

Building on this scope, we formalize the following research question:

\textit{How stably do LLMs maintain assigned student personas across independent conversations and throughout extended conversations?}

To this end, we advance the persona discourse along two methodological directions: a \textit{dual-assessment framework} adapted from clinical psychology that jointly measures self-reported characteristics and observer-rated persona expressions to detect stability failures that are invisible to single-source evaluation \cite{olino_psychometric_2015}; and the joint quantification of between-conversation and within-conversation stability.
We define persona as a specified configuration of characteristics, behavior, and goals that an LLM is instructed to embody.
We ground this investigation in ADHD-related persona specifications: ADHD affects approximately 11\% of school-age children in the United States \cite{danielson_adhd_2024}, making students with varying ADHD symptom profiles a central population for adaptive tutoring and inclusive educational technology.

Two experiments assess two dimensions of temporal stability: 
Experiment~I tests between-conversation stability across the tested models, prompts, and independent runs ($N=4{,}962$, 50 independent runs per condition); 
Experiment~II tests within-conversation stability ($N=3{,}952$, 20 independent conversations consisting of $9$ turns). 

We simulate an educational scenario involving various school settings and use a workplace scenario to test the generalizability of our results.
Furthermore, we vary (1)~model choice, testing five LLMs, and (2)~prompt design, testing three alternative formats intended to represent the same persona-intensity conditions. 

\textbf{Our Contributions:}
\begin{enumerate}[noitemsep, topsep=2pt]
    \item \textbf{Stable within-conversation self-reports, dissociated behavioral expression.}
Within extended conversations, self-reported persona scores remain largely stable, but observer-rated behavioral expression declines during unscripted interactions for high- and moderate-intensity personas. Stability checks based on self-report alone can therefore overstate a simulation's behavioral stability.

    \item \textbf{Interaction structure as the critical moderator.}
We show that lightly scripted conversational structure, anchoring the interaction with recurring task-relevant prompts, nearly eliminates behavioral drift (up to 97\% reduction) without requiring model changes or fine-tuning.

\item \textbf{Intensity-dependent stability.}
Across independent conversations, both self-reports and observer ratings are stable for high-intensity personas but less stable for moderate-intensity personas, indicating a boundary condition.

\item \textbf{A dual-assessment methodology for evaluating human simulations}, adapted from clinical psychology, that jointly measures self-reported and observer-rated persona expression. This method detects self-observer dissociation that single-source (self-report-only or observer-only) evaluation protocols often miss.

\end{enumerate}

\section{Background and Related Work}
\label{sec:related-work}

\subsection{LLM-Based Student Simulation and Pedagogical Applications}

Evidence on LLM-based student simulation is mixed: emerging work shows both potential and clear limitations.
On the promising side, \citet{wu_embracing_2025} showed that LLM agents can simulate students at diverse cognitive levels by grounding persona representations in knowledge graphs built from prior learning records, which substantially improved simulation accuracy and realism.
Similarly, \citet{xu_classroom_simulacra_2025} combined course-material context with longitudinal student interaction data to simulate dynamic learning behavior.
Moving from learner modeling to pedagogical applications, \citet{jin_teachtune_2025} used simulated students to help teachers evaluate pedagogical agents, while \citet{pan_tutorup_2025} used scenario-based simulated students to train novice tutors to address engagement challenges.
At the classroom interaction level, \citet{zhang_simulating_2025} demonstrated that LLM-empowered multi-agent systems can simulate dynamic teacher-student and student-student interactions.
However, research also shows significant limitations.
\citet{martynova_can_2025} conducted semi-structured interviews with teachers who had tutored LLM-simulated students and documented systematic authenticity misalignments, such as the fact that simulated students exhibited uniformly high language complexity, which is inconsistent with genuine learner profiles, and lacked emotional responsiveness.
\citet{li_can_2025} provided large-scale empirical evidence that LLMs systematically misalign with human learners on item difficulty.
Although the models were instructed to simulate the cognitive challenges of students at specific proficiency levels, they still performed well on tasks.
These findings suggest that current LLM architectures may struggle to display the full spectrum of human learner characteristics, such as low performance or nuanced emotional responsiveness. 
Sycophancy compounds this problem in multi-turn educational dialogs.
Models favor responses toward socially desirable or contextually expected outputs rather than maintaining assigned positions \cite{sharma_towards_2025}, a tendency that directly undermines the stability of non-normative student profiles.
Persistent interaction histories and memory profiles can amplify agreement sycophancy, underscoring the need to evaluate persona stability in extended rather than zero-shot interactions \cite{jain_interaction_context_2026}.
A simulated student with high ADHD symptom intensity, for instance, may gradually produce more focused, organized, and compliant responses as the conversation progresses, not because of genuine learning, but because such responses are more consistent with the model's reward-trained priors. 

\subsection{Temporal Persona Consistency and Its Evaluation}

Current evaluation paradigms capture only some stability aspects: most research evaluating human simulation treats model output as a single-trial result and implicitly assumes reproducibility without testing this assumption or examining design decisions such as model choice and prompting strategy \cite{cummins_threat_2025}.
Limited evaluation and model diversity in GenAI persona research further motivate systematic validation across models and assessment approaches \cite{amin_personas_2026}.
This limitation also applies to closely related educational systems: evaluations establish profile alignment, learning-behavior fidelity, or practical utility, but do not test whether assigned learner characteristics remain stable across repeated or extended conversations \cite{jin_teachtune_2025,xu_classroom_simulacra_2025,pan_tutorup_2025}.
Benchmarks assess persona stability in educational or dialog contexts \cite{abdulhai_consistently_2025}, yet research often targets generic user-simulation roles rather than diagnostic-framework-informed profiles, whose coherence is defined by established diagnostic constructs and can be measured with validated instruments.
Critically, reliance on single-source assessments cannot detect dissociations between a model's internal persona representation and its behavioral expression.
Multi-method assessment, jointly measuring self-reports and behavioral indicators, is therefore essential for valid persona evaluation \cite{olino_psychometric_2015}.
Prior work has documented persona drift over extended interactions \cite{gonnermann-muller_maintaining_2026}.

Limited work has directly examined temporal persona consistency.
Prior studies found internally coherent value profiles within individual sessions \cite{rozen_llms_2024}, while consistency-oriented fine-tuning reduced contradictions in multi-turn interactions by more than 55\% \cite{abdulhai_consistently_2025}.
However, it remains unclear whether personas informed by diagnostic frameworks remain stable both across independent conversations and throughout extended interactions, or whether self-reported persona characteristics and observable behavioral expressions exhibit different stability patterns.

\section{Conceptualizing Temporal Persona Stability}
\label{sec:conceptualization}

We conceptualize a persona as an interactional specification: a configuration of characteristics, behaviors, and goals that an LLM is instructed to express in its responses. Temporal persona stability therefore concerns the reproducibility and persistence of that expression under a defined prompting and interaction procedure. It is distinct from representational validity. A persona can be expressed consistently without accurately representing the experiences of the people it purports to simulate, and an authentic human characteristic can be context-dependent rather than uniformly visible in every conversational segment.

We distinguish two temporal dimensions. \textit{Between-conversation stability} describes the variability of persona-aligned measurements across independent model instantiations within a defined experimental scope. It can be assessed within a fixed model--prompt configuration or, as in Experiment~I, in aggregate across tested implementation choices. \textit{Within-conversation stability} describes whether measured persona expression changes as a multi-turn interaction unfolds. These dimensions capture different potential failures: a simulator may be inconsistent across repeated deployments even if each individual conversation is internally stable, or it may begin consistently and drift as conversational context accumulates.

Our dual-assessment framework also distinguishes elicited self-characterization from observable expression. Self-report scores measure how the persona agent characterizes itself when directly prompted with questionnaire items. Observer-report scores measure which persona-relevant behaviors are expressed in generated narratives or conversational segments and are therefore available to an external evaluator. Neither measure provides direct access to a model's internal state. Their convergence or divergence instead characterizes how consistently the assigned persona is elicited through two complementary interaction and measurement procedures.

These conceptual distinctions ground the research question of how stably LLMs maintain assigned student personas across independent conversations and throughout extended conversations.

\section{Study Design}
\label{sec:methodology}

\subsection{Study Overview}

\begin{figure}[t]
    \centering
    \includegraphics[width=\linewidth,alt={Two-panel diagram of the study procedure: Experiment I (left) shows a single-response pipeline of persona prompt, first-person narrative generation, self-report, and three independent observer ratings repeated across 50 runs; Experiment II (right) shows a 9-turn conversation with self-report and observer assessments at turns 3, 6, and 9.}]{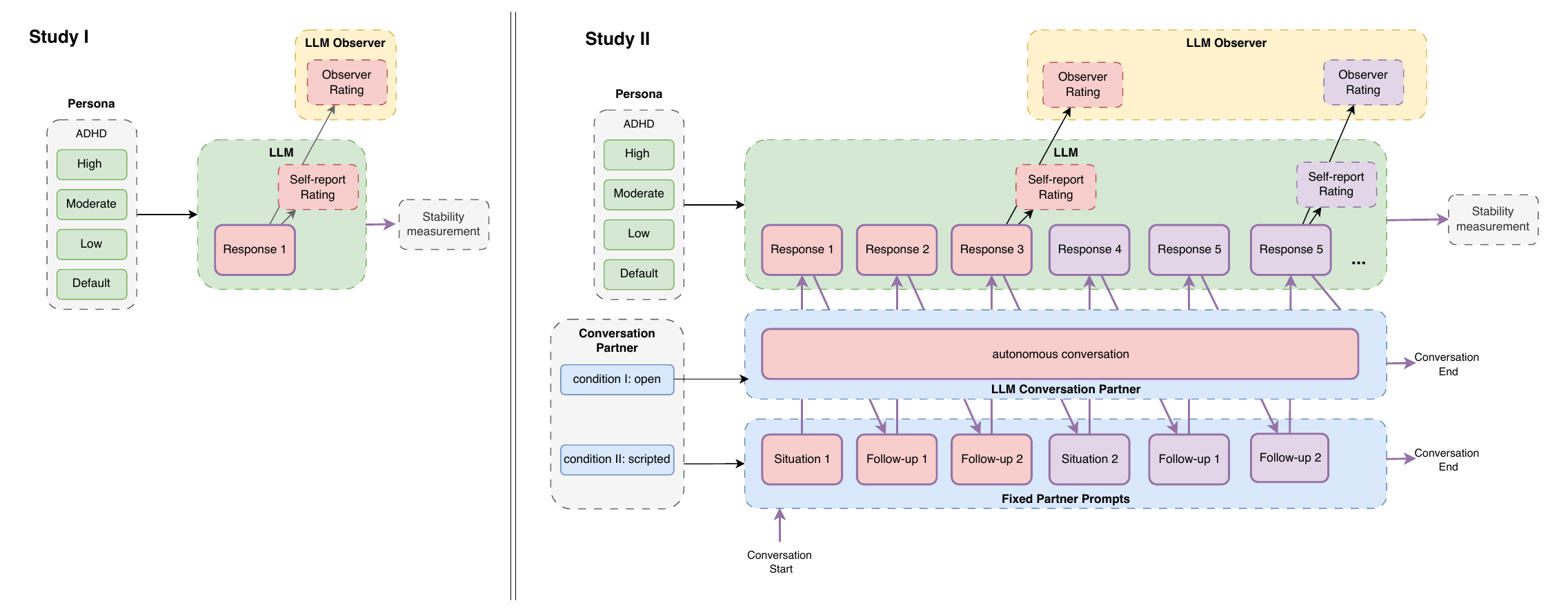}
    \caption{Procedure for Experiment~I (left) and II (right).}
    \label{fig:experiment-pipeline}
    \Description{Two-panel diagram of the study procedure. Experiment I shows a single-response pipeline from persona prompt through narrative generation, self-report, and three independent observer ratings, repeated across 50 runs. Experiment II shows a nine-turn conversation with self-report and observer assessments at turns 3, 6, and 9.}
\end{figure}

We operationalize stability through two complementary experiments (Figure~\ref{fig:experiment-pipeline}). 

The study crosses two temporal dimensions with four persona conditions, three prompt formats, five model families, and educational and workplace scenarios. 
Experiment~I evaluates aggregate stability across independent conversations and tested configurations, whereas Experiment~II evaluates change within multi-turn interactions and contrasts an unscripted condition using adaptive LLM-generated conversation partner responses with a scripted condition using fixed partner-side prompts sent directly to the persona agent.

\subsection{Experimental Factors}

\paragraph{Scenarios and situations.} 

The agent simulation is integrated into two scenarios.
The primary scenario consisted of an educational context.
In experiment~I, a student agent reflects on a typical school day.
In experiment~II, the student agent participates in a conversation with a conversation partner (a friend) (see Appendix ~\ref{app:experiment_taskII_education}).
In the unscripted condition, they have an open conversation about the school day.
In the scripted condition, predefined partner-side prompts ask the student about three school-specific situations, exploring their inner experiences during classroom lectures, collaborative group work, and homework management with standardized follow-up questions (``Can you help me understand that better?'' and ``
What do you mean by that?'') (see Appendix Table~\ref{tab:exp2_situations}). 
To examine the generalizability of the results, a second scenario replicated the same conversational structure in a workplace context, presenting the task of reflecting on a typical workday (see Appendix Table~\ref{app:experiment_taskI_workday}).

\paragraph{Persona specification and experimental intensity.}

The final persona specifications cover the core ADHD-related symptom domains described in the ICD-11 entry for attention deficit hyperactivity disorder (6A05): inattention and hyperactivity/impulsivity \cite{world_health_organization_adhd_2025}. ICD-11 serves as a conceptual framework rather than a diagnostic instrument. The specifications translate these domains into behavioral descriptions and experimentally vary their instructed frequency across low-, moderate-, and high-intensity conditions; the default condition receives no intensity specification. These labels denote experimental prompt intensity rather than clinical severity or diagnosis. The specifications do not operationalize diagnostic thresholds, symptom duration, age of onset, cross-situational presence, or functional impairment.
See all persona prompts in Appendix Table~\ref{tab:persona_prompts_edu}.

\paragraph{Model choice and prompt design.}

We utilized five models spanning proprietary LLMs from different model families: Claude Opus 4.5 (Anthropic), DeepSeek V4 Flash (DeepSeek), GPT 5.1, Gemini 3.1 Pro (Google), and Grok 4.1 Fast (xAI).
This selection encompasses leading proprietary systems from major US and Chinese AI firms, providing comprehensive coverage of current large language model capabilities.
Proprietary models were queried through official provider APIs.
All model queries used provider-default decoding parameters (temperature, top-p, and top-k) to reflect authentic user-facing interactions and avoid confounding variation introduced by custom sampling strategies.
We use three alternative prompt formats intended to operationalize the same persona-intensity conditions: a \textit{text-based prompt} expressing characteristics through frequency adverbs; a \textit{scale-based prompt} encoding characteristics using $7$-point ratings \cite{jebb_review_2021}; and a \textit{paraphrased prompt} varying wording and information order (see Appendix Table~\ref{tab:persona_prompts_edu}).

\subsection{Experiment I: Between-Conversation Procedure}

Each of the $4{,}962$ conversations was a fully independent, one-shot instantiation with no shared context between runs.
The procedure per run was: (1) the target LLM received a persona prompt in one of three formats (text-based, scale-based, or paraphrased) specifying a high, moderate, or low ADHD intensity level. 
(2) It generated a first-person narrative of a typical school day (or workday in the generalization condition) in response to a scenario-specific task prompt (see Appendix~\ref{app:experiment_taskI_education}). The high-, moderate-, and low-intensity conditions received the ADHD-explicit narrative task, whereas the default condition received a neutral task that asked only for a description of a typical schoolday or workday. 
(3) Immediately after, the same LLM completed the 26-item CAARS self-report short form. We extracted its 12-item ADHD Index, yielding scores from $0$ to $36$.
(4) Three independent LLM raters, blind to persona instructions, evaluated the narrative using the 30-item CAARS observer-report short form. We likewise extracted the 12-item ADHD Index from each response. 
This procedure was repeated 50 times per condition, with each run starting from a clean context to ensure statistical independence.

\subsection{Experiment II: Within-Conversation Procedure}

Each of the $3{,}952$ conversations was a multi-turn interaction between a persona agent and a neutral conversation partner.
The procedure per conversation was: 
(1) The persona agent received the same persona prompt as in Experiment~I. 
(2) In the unscripted condition, the conversation partner (Gemini 3.1 Pro) was instantiated as a neutral listener instructed to ask follow-up questions without sharing opinions or influencing the persona agent's responses. No conversation-partner model was used in the scripted condition (see Appendix~\ref{app:conversationpartner_unscripted}).
(3) The agents engaged in a conversation for $9$ turns, with stability assessments at turns $3, 6,$ and $9$.
At each checkpoint, the persona agent completed the CAARS self-report questionnaire, and three independent LLM raters evaluated the preceding $3$-turn segment using the observer-report form, blind to persona instructions. The CAARS assessment was administered in a separate branch from a snapshot of the dialogue history; neither the questionnaire nor its response was retained in the ongoing conversation context, which continued from the pre-assessment history.
Here, one conversation turn consisted of one response from the conversation partner followed by one response from the persona agent. Thus, a $9$-turn conversation contained $18$ responses in total. At each checkpoint, the self-report captured the persona agent's current characterization of itself, while each observer independently rated only the three conversational turns since the previous checkpoint. This produced three comparable assessment segments covering turns $1$--$3$, $4$--$6$, and $7$--$9$.
We varied two conversation conditions: In the \textit{unscripted} condition, the persona agent and conversation partner engaged freely, with the partner asking open-ended follow-up questions about the school day (or workday). 
In the \textit{scripted} condition, nine predefined partner-side prompts were sent directly to the persona agent. 
They covered three pre-specified situations in randomized order, each followed by two fixed follow-up questions. 
In both conditions, the persona agent's responses were generated freely.
Full situation prompts are provided in Appendix Table~\ref{tab:exp2_situations}.

\subsection{Measures and LLM Observer Ratings}

We use the Conners' Adult ADHD Rating Scales (CAARS) \cite{conners_conners_2012}, a validated instrument with parallel self-report and observer-report forms.
Both outcomes are the $12$-item ADHD Index (range: $0-36$). The full $26$-item self-report and $30$-item observer short forms were administered; index items were then extracted from the responses (see Appendix Section~\ref{sec:ratingprompts}). 
Three independent LLM raters (Claude Opus 4.5, GPT-5.1, Gemini 3.1 Pro) rated each narrative independently.
We selected three instruction-following LLMs from different providers to reduce reliance on a single model family.
Observer-report scores were aggregated across three independent LLM evaluators, a decision justified by high Intraclass Correlation Coefficients (ICC). 
We assessed inter-rater reliability using ICC (ICC(2,1)).
In Experiment~I, reliability was high for the education condition (ICC(2,1) = $.92$, $95\%$ CI $[.76, .96]$). 
In Experiment~II, the unscripted conditions showed moderate to good reliability for the education condition (ICC(2,1) = $.72$, $95\%$ CI $[.25, .87]$).
The scripted conditions demonstrated substantially higher reliability (ICC(2,1) = $.87$, $95\%$ CI $[.53, .94]$) (see Appendix for ICC overview Table~\ref{tab:icc_results}).

\subsection{Human-Rater Validation}

We assessed human--LLM rating convergence by comparing LLM observer ratings with human expert judgments on $20$ narratives randomly sampled from the Experiment~I workday dataset. Five M.Sc.-level psychologists, blind to the persona condition, each independently rated all $20$ narratives.
Human inter-rater reliability was excellent (ICC(2,1) = $.92$, $95\%$ CI $[.85, .96]$), LLM reliability was good (ICC(2,1) = $.87$, $95\%$ CI $[.55, .95]$), and human--LLM rating convergence was excellent (ICC(2,1) = $.95$, $95\%$ CI $[.57, .99]$).

This expert-rating comparison validates agreement on the application of the observer instrument to generated text; it does not validate the personas against the lived experiences or behavior of people with ADHD. The remaining experiments involve generated model outputs rather than interactions with students, teachers, or people represented by the personas.

\subsection{Data Accounting and Statistical Analysis}

We compute independent ADHD intensity ratings from both self-report and observer-report forms; higher values indicate greater ADHD symptom intensity.
Experiment~I comprised $N = 4{,}962$ one-shot conversations (5 models $\times$ 3 personas $\times$ 3 prompts $\times$ 2 scenarios $\times$ 50 runs + 5 models $\times$ 2 scenarios $\times$ 50 runs with default configuration) with $N=14{,}886$ observer assessments and $N=4{,}962$ self-reports.
Experiment~II comprised $N = 3{,}952$ multi-turn conversations (5 models $\times$ 3 personas $\times$ 3 prompts $\times$ 2 conditions $\times$ 2 scenarios $\times$ 20 runs + 5 models $\times$ 2 conditions $\times$ 2 scenarios $\times$ 20 runs with default configuration).
Each conversation extended to 9 turns (18 responses) with assessments at three checkpoints, yielding $N = 35{,}430$ observer assessments (turn 3: $n = 11{,}799$; turn 6: $n = 11{,}808$; turn 9: $n = 11{,}823$) and $N = 11{,}796$ self-report measurements (turn 3: $n = 3{,}921$; turn 6: $n = 3{,}923$; turn 9: $n = 3{,}952$). 
In Experiment~I, $4{,}962$ of $5{,}000$ targeted conversations ($99.24\%$) entered the analysis; in Experiment~II, $3{,}952$ of $4{,}000$ targeted conversations ($98.80\%$) entered the analysis with at least one paired observer and self-report assessment. The remaining runs ($0.76\%$ and $1.20\%$ of targeted conversations, respectively) produced no usable paired record due to API connectivity interruptions or malformed JSON outputs of questionnaire items that prevented automated parsing (see Appendix Table~\ref{tab:cell_counts} in the Appendix). In addition, a small number of individual assessments failed even in conversations that completed, so checkpoint-level missingness is not monotone.

For Experiment~I, we operationalize aggregate between-conversation stability as the standard deviation ($SD$) of ADHD Index scores within each persona and scenario after pooling across the tested models, prompt formats, and independent runs, with higher variability indicating lower stability. This metric captures both run-to-run variation and sensitivity to model and prompt choice; it therefore describes stability across the tested implementation space rather than repeatability with model and prompt held fixed. SD is also reported as a percentage of the scale range, calculated as $SD / \text{scale range} \times 100$.

We conducted separate analyses for self-report and observer-report measures.
Additionally, we conducted linear mixed-effects models (LMMs) using the \textit{lme4} and \textit{lmerTest} packages to examine the stability of ADHD persona simulations across conversations.
For each scenario, interaction structure (unscripted and scripted), and assessment type (self-report and observer-rated), we modeled the respective scale score as a function of assessment checkpoint, persona intensity, prompt, and model, with random conversation-level intercepts to account for repeated assessments within conversations; these separate models did not include random checkpoint slopes.
Checkpoint was entered as a numeric predictor coded $1,2,3$, corresponding to turns $3,6,9$; coefficients therefore represent change per successive three-turn assessment interval, and fitted first-to-last change equals twice the checkpoint coefficient.
Default-persona runs were excluded because they have no assigned persona intensity to compare against. The education models included $898$ unscripted and $900$ scripted conversations; the workday models included $884$ unscripted and $898$ scripted conversations.
Post-hoc pairwise comparisons among checkpoints used estimated marginal means with Bonferroni correction.
The results presentation in section~\ref{sec:results} focuses primarily on education-related conditions, with complete fixed-effect results for the workday conditions in the Appendix.

\section{Results}
\label{sec:results}

We report the results in the order of the experiments. Experiment~I addresses aggregate between-conversation stability across tested configurations. Experiment~II addresses within-conversation change and the divergence between assessment modes, while comparisons across interaction structures, models, prompts, and scenarios characterize the moderators of these patterns.

\subsection{Between-Conversation Stability (Experiment~I)}

\begin{figure}[t]
    \centering
    \begin{subfigure}[t]{0.49\textwidth}
        \centering
        \includegraphics[width=\linewidth,alt={Box plots of pooled self-reported ADHD Index scores in Experiment I, by persona intensity (default, low, moderate, high) for the education and workday scenarios, aggregating tested models, prompt formats, and independent runs.}]{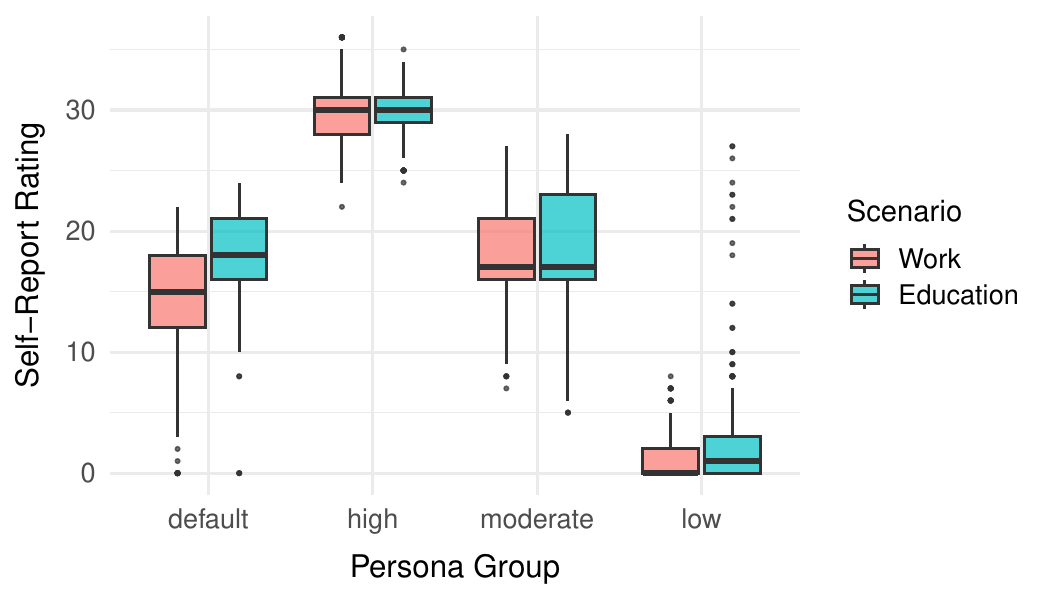}
        \subcaption{}
        \label{fig:self-report_stability_exp1}
    \end{subfigure}
    \hfill
    \begin{subfigure}[t]{0.49\textwidth}
        \centering
        \includegraphics[width=\linewidth,alt={Box plots of pooled observer-rated ADHD Index scores in Experiment I, by persona intensity (default, low, moderate, high) for the education and workday scenarios, aggregating tested models, prompt formats, and independent runs.}]{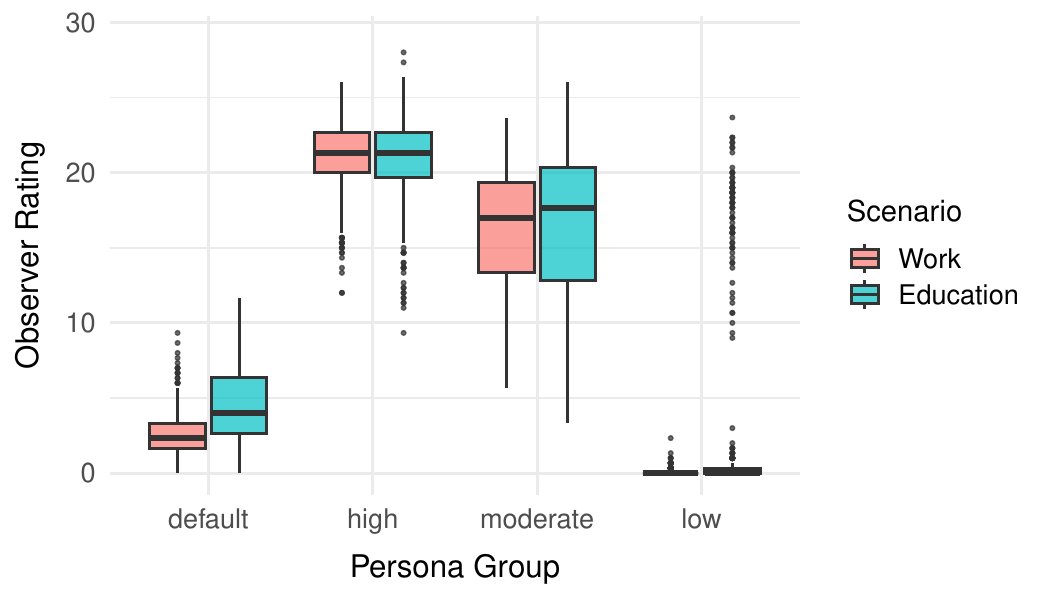}
        \subcaption{}
        \label{fig:observer_stability_exp1}
    \end{subfigure}
    \caption{Experiment~I: Pooled ADHD Index distributions by persona intensity and scenario across the tested models, prompt formats, and independent runs. (\subref{fig:self-report_stability_exp1}) Self-report; (\subref{fig:observer_stability_exp1}) observer scores averaged across three LLM raters per conversation before pooling.}
    \label{fig:stability_exp1}
    \Description{Two sets of box plots show pooled self-reported and observer-rated ADHD Index scores across independent runs and tested model and prompt configurations, grouped by default, low-, moderate-, and high-intensity personas in the education and workday scenarios. High-intensity scores cluster near the upper end and low-intensity scores cluster near zero. Default self-reports align with the moderate range, whereas default observer ratings remain near the low end, slightly above the workday default.}
\end{figure}

We first examined aggregate persona stability across independent conversations, models, and prompt formats.
Figure~\ref{fig:self-report_stability_exp1} and Figure~\ref{fig:observer_stability_exp1} present the pooled score distributions used to assess self-reported and observer-rated stability across persona intensity levels in the education scenario (see Appendix Table~\ref{tab:exp1_descriptives} for descriptive statistics of the education and workday scenarios).

\paragraph{Stability by persona intensity and assessment mode.}

Self-reported scores (scale: $0-36$) demonstrated an intensity-dependent pattern in the education scenario: high-intensity scores had lower aggregate variability, whereas moderate-intensity scores had higher aggregate variability.
High-intensity personas produced tight clustering across the pooled observations (education: $M = 30.2$, $SD = 1.79$). For context, a sample of 43 clinically stable adults with ADHD scored $11.5$ (SD = $4.0$) on the ASRS Screener \cite{matza_testretest_2011}.
Low-intensity personas had a pooled self-report mean of $2.08$ ($SD = 3.35$, $9.3\%$ of the scale) in the education scenario.
However, the elevated SD in the low condition is largely driven by outliers, as shown in Figure~\ref{fig:stability_exp1}.
Moderate-intensity personas showed higher aggregate self-report variability in the education scenario ($M = 18.0$, $SD = 5.04$, $14.0\%$). This pooled estimate combines variation across repeated runs, prompt formats, and model families.
Models, therefore, lack stable representational attractors for this intermediate-intensity region.

Observer-rated scores (scale: $0-36$) showed a similar intensity-dependent pattern.
High-intensity personas had lower aggregate variability in the education scenario ($M = 21.1$, $SD = 2.57$, $7.1\%$).
Moderate-intensity conditions showed higher aggregate variability ($M = 16.6$, $SD = 4.78$, $13.3\%$).
Low-intensity observer ratings showed the highest pooled variability ($M = 1.71$, $SD = 5.07$, $14.1\%$), reflecting combined variation across runs, prompts, and models.
Again, this is mainly driven by extreme outliers. 

Comparing pooled self-report and observer-rating means revealed descriptive differences between assessment modes, with self-report means generally exceeding observer-rating means across persona conditions and scenarios.

\paragraph{Default-persona scores.}

Default personas received only the minimal persona instruction ``You are a student'' and a neutral task prompt, providing a baseline under the student and schoolday framing.
Default self-reports clustered in the moderate range ($M = 18.2$, $SD = 3.61$), matching the moderate-intensity mean ($M = 18.0$) and remaining well above the low-intensity target ($M = 2.08$).
Observer-rated default behavior was much lower ($M_{\text{obs}} = 4.44$, $SD = 2.60$), closer to the low-intensity mean ($M = 1.71$) than to moderate ($M = 16.6$) or high ($M = 21.1$).
In the education scenario, default personas produced self-report scores aligned with the moderate-intensity condition, while their observer ratings remained much closer to the low-intensity condition.
This self-report--observer gap also appeared in the workplace scenario, where default employee personas scored $M = 13.9$ ($SD = 4.91$) on self-report and $M_{\text{obs}} = 2.78$ ($SD = 1.68$) on observer ratings.
Both measures were lower in the workplace than in the education scenario (see Figure~\ref{fig:self-report_stability_exp1} and Figure~\ref{fig:observer_stability_exp1}).

\paragraph{Generalizability and influence of additional factors.}

The workplace scenario largely replicated the educational findings (Table~\ref{tab:exp1_descriptives}).
Descriptive data (Table~\ref{tab:exp1_persona_prompt}) reveal consistent prompt-level differences.
For moderate-intensity workday personas, the text-based prompt produced substantially higher self-report means and observer means than the scale-based prompt, indicating that qualitative frequency-adverb descriptions elicit stronger ADHD expression than quantitative Likert-encoded specifications.
The paraphrased prompt yielded intermediate values.
Model-level analysis (Table~\ref{tab:exp1_persona_model}) revealed substantial heterogeneity, particularly for moderate and default personas.
DeepSeek models produced higher observer ratings for moderate-intensity workday personas, whereas Claude produced notably lower ratings for the same condition.
These model-level differences confirm that the model family substantially moderates the level of persona expression.

\subsection{Within-Conversation Stability (Experiment~II)}

\begin{figure*}[t]
    \centering
    \begin{subfigure}[t]{0.49\textwidth}
        \centering
        \includegraphics[width=\linewidth,alt={Line plot of mean self-reported ADHD Index across conversation turns 3, 6, and 9 by persona intensity (high, moderate, low, default) with error bars, showing stable self-reports across turns.}]{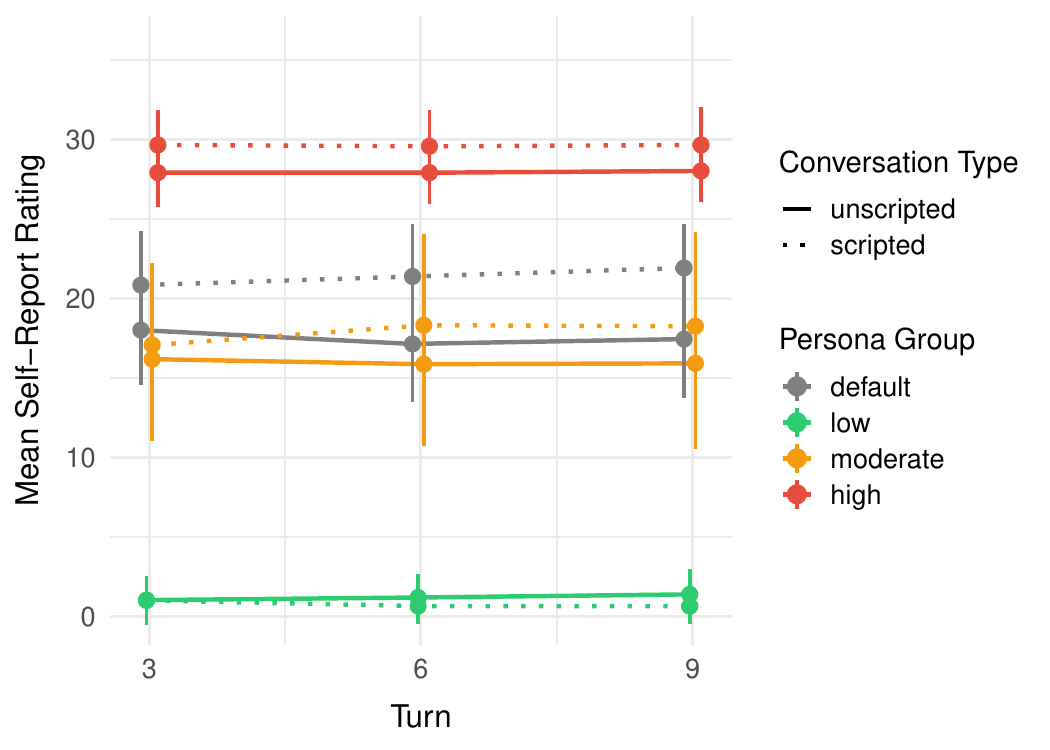}
        \subcaption{}
        \label{fig:self-report_stability_exp2}
    \end{subfigure}
    \hfill
    \begin{subfigure}[t]{0.49\textwidth}
        \centering
        \includegraphics[width=\linewidth,alt={Line plot of mean observer-rated ADHD Index across conversation turns 3, 6, and 9 by persona intensity (high, moderate, low, default) with error bars, showing declining observer ratings for high- and moderate-intensity personas within unscripted conversations.}]{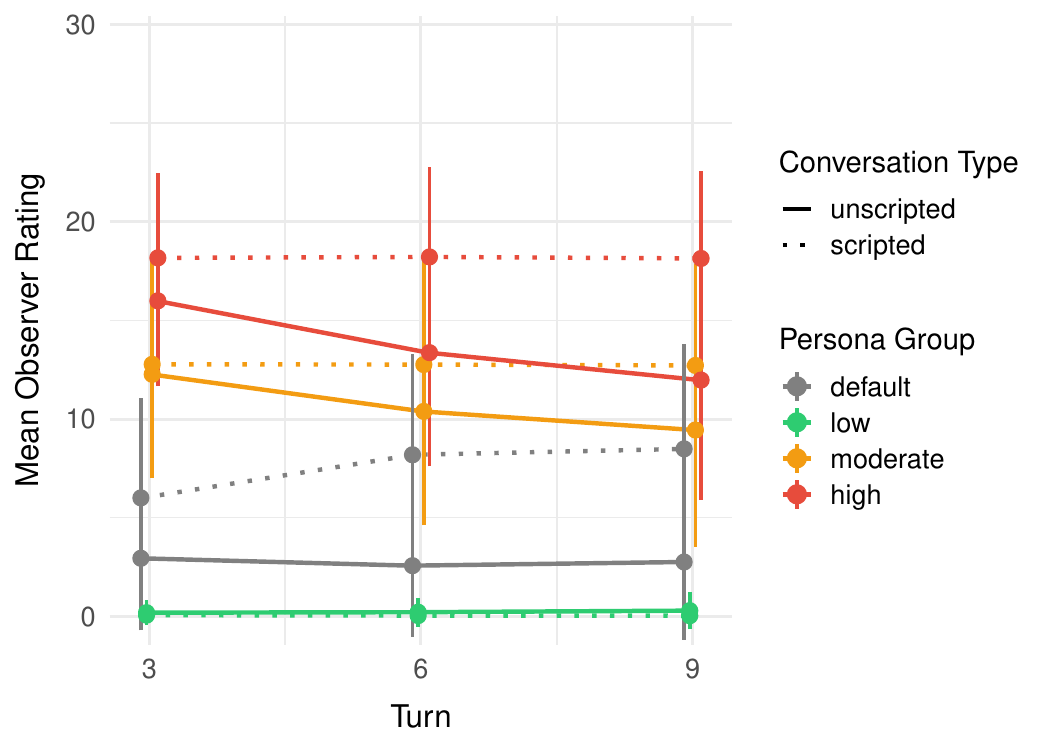}
        \subcaption{}
        \label{fig:observer_stability_exp2}
    \end{subfigure}
    \caption{Mean ADHD Index in Experiment~II for education conversations by persona across prompt groups, with error bars indicating $\pm SD$ (high=red, moderate=yellow, low=green, default=grey). (\subref{fig:self-report_stability_exp2}) Self-report; (\subref{fig:observer_stability_exp2}) observer ratings.}
    \Description{Two line plots show mean ADHD Index scores with standard-deviation error bars at conversation turns 3, 6, and 9 for high-, moderate-, low-intensity, and default personas. Self-reported scores remain approximately stable, while observer ratings decline across turns for high- and moderate-intensity personas in unscripted conversations.}
\end{figure*}

Figure~\ref{fig:self-report_stability_exp2} and Figure~\ref{fig:observer_stability_exp2} present self-reported and observer-rated stability measures across persona intensity levels in the educational scenario.

\begin{table}[t]
    \centering
    \caption{Turn effects from the linear mixed models for within-conversation stability in the education scenario.
    Entries report the fixed effect of assessment checkpoint. Models additionally include persona intensity, prompt design, and LLM, with random intercepts for conversations. Default personas are excluded from all models. Full model results are reported in Appendix Tables~\ref{tab:unscripted_sr_edu}--\ref{tab:scripted_obs_edu}.}
    \label{tab:exp2_lmm_turn_effects}
        \begin{tabular}{llrrrr}
        \toprule
        \textbf{Assessment} & \textbf{Interaction} & $\boldsymbol{\beta}$ & \textbf{SE} & $\boldsymbol{t}$ & $\boldsymbol{p}$ \\
        \midrule
        Self-report & Unscripted &  0.03 & 0.04 &  0.90 & .366 \\
        Self-report & Scripted   &  0.14 & 0.04 &  3.27 & .001 \\
        Observer    & Unscripted & $-1.13$ & 0.05 & $-20.80$ & $<.001$ \\
        Observer    & Scripted   & $-0.02$ & 0.05 & $-0.36$ & .722 \\
        \bottomrule
    \end{tabular}

    \end{table}

\paragraph{Stable self-reported within-conversation stability.}

Self-reports ($0-36$) remained stable throughout extended conversations across all persona conditions and interaction structures in the educational scenario. 
For high-intensity education personas, drift across the full conversation was negligible: unscripted $\Delta = 0.1$ points ($0.3\%$), scripted $\Delta = 0.0$.
Moderate-intensity personas showed similarly minimal self-report change: unscripted education $\Delta = -0.3$ ($0.8\%$ of scale), scripted education $\Delta = 1.2$ ($3.3\%$ of scale).
Low-intensity personas exhibited slightly higher absolute drift (unscripted $\Delta = 0.36$, $1.0\%$ of scale; scripted $\Delta = -0.36$, $1.0\%$ of scale), though all changes remained within measurement error. 
A formal LMM analysis confirmed this pattern (Table~\ref{tab:exp2_lmm_turn_effects}).
In the unscripted education condition, the effect of turn number on self-reported ADHD scores was non-significant ($\beta = 0.03$, $SE = 0.04$, $t = 0.90$, $p = .366$), indicating no statistically detectable drift across turns $3, 6,$ and $9$.
Although no drift is apparent in the scripted education condition, the scripted self-report model showed a small but statistically significant positive trend.
Given the scripted sample size ($N = 2{,}700$ observations), this effect is statistically detectable but practically small ($\beta = 0.14$, $SE = 0.04$, $t = 3.27$, $p = .001$; Table~\ref{tab:exp2_lmm_turn_effects}).
Thus, education self-reports were stable or changed only modestly on average across the tested interaction structures. This relative stability provides evidence that models can maintain the assigned self-characterization during an interaction, while the small scripted trend cautions against interpreting it as perfect invariance.

\paragraph{Scripted vs. unscripted within-conversation effects.}

Observer ratings ($0-36$) revealed a critical divergence from the self-report pattern: a progressive decline in externally visible persona expression across conversation turns, but only in unscripted interactions. 
In the unscripted education condition, high-intensity personas declined by $\Delta = -4.0$ points ($11.1\%$).
Moderate-intensity personas showed a parallel decline ($\Delta = -2.82$, $7.8\%$).
Low-intensity personas showed minimal drift ($\Delta = 0.10$, $0.3\%$), consistent with the near-floor absolute ratings, leaving little room for further reduction.
Interestingly, scripted conversations showed near-complete elimination of this drift.
High-intensity scripted education personas showed $\Delta = -0.1$ ($0.3\%$) across the full conversation.
Moderate-intensity scripted personas similarly maintained stability ($\Delta = -0.1$, $0.3\%$).
The contrast between unscripted ($\Delta = -4.0$, $11.1\%$) and scripted ($\Delta = -0.1$, $0.3\%$) high-intensity education observer ratings is a $97\%$ reduction in behavioral drift from conversation structure alone.
The LMM analysis confirmed these patterns (Table~\ref{tab:exp2_lmm_turn_effects}): For unscripted education observer ratings, the turn effect was highly significant and substantial ($\beta = -1.13$, $SE = 0.05$, $t = -20.80$, $p < .001$).
Post-hoc pairwise comparisons confirmed significant differences between all turn pairs (all $p < .001$).
The scripted education observer model yielded a non-significant turn effect ($\beta = -0.02$, $SE = 0.05$, $t = -0.36$, $p = .722$; Table~\ref{tab:exp2_lmm_turn_effects}), with all post-hoc comparisons yielding $p = 1.000$.
Within the separate education models, scripted conversations therefore showed no detectable observer-rated decline, in contrast to the decline observed in the unscripted condition.

\paragraph{Workplace scenario.}
The separate workday models showed the same qualitative pattern (see Appendix section~\ref{sec:ExpIIaddresultsWork} and Tables~\ref{tab:unscripted_sr_work}--\ref{tab:scripted_obs_work}).
For self-reports, the unscripted workday model showed no detectable checkpoint trend ($\beta = 0.04$, $SE = 0.04$, $t = 1.01$, $p = .311$), whereas the scripted model showed a small positive trend ($\beta = 0.27$, $SE = 0.04$, $t = 6.31$, $p < .001$), corresponding to a fitted increase of $0.53$ points ($1.5\%$ of the scale range) from the first to the last checkpoint.
For observer ratings, the unscripted workday model showed a significant checkpoint decline ($\beta = -0.82$, $SE = 0.05$, $t = -17.63$, $p < .001$; all Bonferroni-adjusted checkpoint comparisons $p < .001$), whereas the scripted model showed no detectable checkpoint effect ($\beta = -0.03$, $SE = 0.04$, $t = -0.71$, $p = .477$; all pairwise $p = 1.000$).
Because the scripted and unscripted models were fitted separately, these results establish the within-condition trajectories but do not directly test whether the two checkpoint slopes differ.

\paragraph{Model-level influences.}
The LMM results across education conditions (Tables~\ref{tab:unscripted_sr_edu}--\ref{tab:scripted_obs_edu}) reveal consistent model-level differences.
In the scripted self-report model, DeepSeek and Grok produced higher scores than Claude, while GPT 5.1 produced lower scores.
Gemini showed a smaller, non-significant deviation.

\section{Discussion} 
\label{sec:discussion}

We examined whether LLMs maintain stable persona representations across and within conversations.
We demonstrate three core findings: (1) LLMs produce comparatively stable, persona-aligned self-reports across both between-conversation and within-conversation settings for high personas; (2) observer-rated behavioral expression drifts systematically in unscripted conversations for high- and moderate-intensity personas; and (3) scripted conversations with explicit task-relevant prompts show little observer-rated drift for these personas, including a $97\%$ descriptive reduction for high-intensity education personas.

\subsection{Stable Self-Reports and the Self–Observer Dissociation}

The comparatively high aggregate self-report stability of high-intensity personas and the limited within-conversation change are not trivial results.
Our persona prompts consisted of brief text descriptions or aggregated scale values that do not directly map onto specific questionnaire items.
Producing stable responses requires the model to translate abstract specifications, such as ``frequently struggles to maintain attention'', into coherent responses across psychologically distinct items (such as ``I avoid new challenges because I lack faith in my abilities'').
If this were simple surface-level instruction-following, we would expect inconsistency across items that lack lexical overlap with the persona prompt. 
The stability across varied item formulations instead suggests that LLMs construct and maintain personality representations that generalize appropriately across diverse item content.
The moderate-intensity exception highlights a critical boundary, but it points to a training data gap rather than an architectural limitation.
Training data contain clinical case material for diagnosed ADHD and normative developmental narratives, but rarely feature the threshold-straddling presentations that define moderate symptom expression.
Models, therefore, lack stable representational attractors in this intermediate region. 
An alternative interpretation might be that perfectly stable moderate ADHD is psychologically unrealistic.
The elevated variability of mildly expressed ADHD personas may not solely reflect representational deficiencies in LLMs, but may instead mirror the intrinsic heterogeneity of ADHD itself.
ADHD, particularly subclinical presentations, is widely described as behaviorally heterogeneous and context-dependent rather than a single stable profile \cite{wahlstedt_heterogeneity_2009}.

This stability of self-reports, however, contrasts with how the assigned persona manifests in observable behavior.
The central theoretical contribution of our dual-assessment framework is the dissociation it exposes: LLMs reliably report the assigned persona when directly asked, yet their spontaneous behavioral expression of that persona degrades in unscripted conversation.
These two measurement modes capture fundamentally different aspects of persona simulation.
The direction of the descriptive difference aligns with prior human research showing higher self-ratings than observer ratings for hyperactivity/restlessness and impulsivity symptoms \cite{morstedt_attention-deficithyperactivity_2015}.
Self-report measures whether the LLM can access and accurately characterize its assigned persona when prompted.
Observer rating measures whether the LLM produces conversational content that an external rater would identify as consistent with the specified persona at the specified intensity. 
The two measures therefore provide complementary rather than interchangeable evidence. A stable self-report accompanied by a lower observer score does not necessarily mean that the assigned persona has been forgotten. It may instead indicate that fewer observable or rateable expressions of the persona appear in the conversation. This distinction matters in educational simulation because teachers and tutors primarily encounter the learner's expressed conversational behavior rather than its questionnaire-elicited self-characterization.
This dissociation is invisible to single-source evaluation approaches, which is precisely why the dual-assessment design matters.
Our results show that these measures can diverge substantially, with direct practical consequences.
A teacher or tutor interacting with a simulated student forms judgments based on observable behavior rather than on the student's self-descriptions.
If the simulation produces stable personas that manifest inconsistently in behavioral outputs, the resulting training environment is systematically misleading.

An important alternative interpretation is that declining observer scores reflect a decline in opportunities for rateable persona expression rather than loss of the assigned persona. As an unscripted conversation develops, later segments may move away from situations that elicit descriptions of attention, organization, or impulsivity. The observer instrument can only score behaviors visible in a segment, whereas the self-report directly asks about those characteristics at every checkpoint. Accordingly, the observed divergence should be interpreted as a difference between elicitation and expression under the tested interaction conditions, not as direct evidence that an internal persona was forgotten.

\subsection{Interaction Structure as a Design Variable}

Taken together, these findings position interaction structure as part of the measurement setup rather than as a neutral delivery format. 
By standardizing opportunities for persona-relevant expression at each checkpoint, the scripted condition yielded more temporally consistent observer ratings in these experiments. 
This does not establish that scripting preserved an internal persona state or reveal the underlying mechanism.

This increased consistency introduces a design trade-off. 
Repeated situation-specific prompts make persona-relevant behavior more consistently observable, but a tightly scripted conversation can also become less natural and can mechanically re-elicit the characteristics being measured. 
Higher measured stability therefore does not by itself establish higher simulation fidelity. 
Designers must determine how much structure is needed to create comparable opportunities for expression without scripting the target behavior so strongly that the interaction ceases to represent plausible dialog.

\subsection{Implications for LLM-based Simulation}
\label{sec:implications}

\textit{LLM-based persona simulation is viable but only under specific conditions.} Our findings suggest that stability is conditional on persona intensity, measurement mode, and interaction design. Practitioners should treat these as the boundary conditions for reliable use rather than assume stability holds uniformly. 
HCI researchers adopting LLMs for human simulation should report the specific conditions under which stability they were established and investigate boundary conditions beyond model choice and prompt design, such as interaction structure, that influence simulation stability, rather than treating model selection as the primary lever for reliability.

\textit{Design the interaction structure.} Across five models and two scenarios, structured conversation design produced more temporally consistent observer-rated behavior. Practitioners building simulation-based tools should treat interaction protocols as a primary design variable, not an afterthought to model selection.

\textit{Partial and subclinical profiles may be the hardest to simulate reliably.} Moderate-intensity personas showed the least stable self-reports and observer ratings across our conditions. We raise this as a hypothesis: if the pattern generalizes beyond our ADHD case, it would suggest that current LLM-based simulations are least reliable for exactly the partial, borderline profiles most common in real populations. HCI researchers simulating psychological or behavioral characteristics should test whether stability varies by characteristic intensity, rather than assuming a single validation at one intensity level generalizes across the full range a persona might exhibit.

\textit{Default personas show a self-report--behavior gap.} Under the minimal ``You are a student'' persona instruction and a neutral narrative task, default personas endorsed ADHD characteristics in the moderate range on self-report but displayed only weakly elevated ADHD-like behavior to observers. The same directional pattern appeared at lower levels in the workplace scenario. Designers should therefore assess both elicited self-characterization and observable expression rather than treating a default persona as equivalent to a low-intensity condition or inferring its behavior from self-report alone.

\subsection{Ethical Considerations and Responsible Use}

LLM-based persona simulation offers potential benefits from an ethical perspective.
It may reduce demands placed on participants with specific neurodevelopmental profiles, enable controlled and reproducible preliminary experiments, and help identify ineffective or potentially harmful interaction designs before they are evaluated with real learners.
Simulations may likewise support preliminary exploration of teacher-training scenarios.
These benefits depend, however, on using simulation as a complement to, rather than a replacement for, research involving people with lived experience.
Temporal stability and representational validity are distinct and equally important requirements: the present study examines whether an assigned persona is expressed consistently, but does not establish whether the generated behavior accurately represents people with ADHD.
A stable simulation may still reproduce stereotypes or distortions encoded in its prompts or model, while its consistency may lend those representations unwarranted credibility.
This risk becomes particularly consequential if simulated learners inform teacher training, intervention design, or educational decisions without validation involving the populations represented.
We therefore view persona simulation as most appropriate for preliminary experimentation, methodological stress testing, and hypothesis generation; consequential applications require separate validation and meaningful involvement of relevant communities.

\subsection{Limitations and Future Research}
\label{sec:limitations}

Our findings are subject to limitations that inform directions for future research.
\textit{Construct generalizability.}
We focused on one diagnostic construct (ADHD); generalization to more complex personas consisting of multiple human characteristics remains for future research.
Future work should systematically investigate stability patterns across a broader range of psychological characteristics and with different conversation partner personas.
Extending to comorbid conditions (e.g., ADHD with anxiety) would strengthen ecological validity for educational simulation.
\textit{Scope of interaction structures.}
While we tested scripted vs. unscripted conditions, other interaction structures (e.g., increasing scaffolding, supportive versus challenging tutors) remain unexplored.
A systematic investigation of interaction design variables could identify minimal intervention strategies for maintaining behavioral stability without fully sacrificing naturalness.
\textit{Measurement validity.}
The CAARS-based measures quantify persona-relevant content elicited by our prompts; they do not establish clinical validity or diagnostic equivalence. CAARS was designed and validated for adults, making it appropriate for the adult employee role in the workday scenario but an imperfect fit for the educational scenario, where the school-based student prompts did not specify an age. Scores in the educational scenario should therefore be interpreted as standardized experimental indicators of ADHD-related content rather than age-valid clinical measurements of school-age learners. Observer ratings also depend on which behaviors become visible within each three-turn segment, and the human-rater validation establishes agreement on scoring generated text rather than correspondence with the lived experiences of people with ADHD. Future work should combine psychometric assessment with behavioral coding developed and validated with relevant stakeholders and should test whether findings persist under measures less dependent on explicit symptom expression.
\textit{Interaction and application validity.}
The educational scenario models a conversation with a friend about a school day rather than an enacted teacher-training or tutoring interaction. The results therefore identify a prerequisite and a design consideration for simulated-learner systems, not evidence of their educational effectiveness. Evaluations with educators, learners, and neurodivergent stakeholders are required before drawing conclusions about training outcomes, realism, or appropriate deployment.

\section{Conclusion}
Our dual-assessment framework reveals a critical distinction: LLMs reliably produce stable, persona-aligned self-reports across and within conversations for high intensities. 
Meanwhile, observable behavioral expressions of high- and moderate-intensity personas systematically decline in unscripted conversations. 
Importantly, scripted conversations showed little observer-rated drift for high- and moderate-intensity personas, suggesting that measured stability is conditional on interaction design rather than intrinsic LLM capability.
Taken together, high-intensity personas are reproduced stably across independent conversations, and low-intensity variability stems largely from rare outlier responses; self-reports remain stable while observer-rated expression declines within extended conversations, and interaction structure, model choice, prompt design, and scenario all moderate these stability patterns, with interaction structure being decisive.
Our findings show that LLMs produce stable, persona-aligned self-reports, while stability in extended interactions depends on structured conversation design.
This holds significant implications for applications requiring sustained, path-dependent learner interactions, such as teacher training, adaptive tutoring, or research on emergent learning dynamics, where the gap between what a simulated student could access and what a teacher observes behaviorally is of extreme importance.

\bibliographystyle{plainnat}
\bibliography{references}

\newpage

\appendix

\section{Appendix}
\label{app:appendix}

\subsection{Methodological Supplements}

\subsubsection{Large Language Models used in the study}
\leavevmode
\begin{table}[H]
\centering
\caption{Large language models used in this study.}
\label{tab:LLMs}
\footnotesize
\begin{tabular}{llll}
\toprule
\textbf{Model} & \textbf{Parameters} & \textbf{Provider} & \textbf{Role} \\
\midrule
Claude Opus 4.5& Undisclosed & Anthropic & Persona + Evaluator \\
DeepSeek V4 Flash& 284B total, 13B active & DeepSeek & Persona \\
GPT 5.1 & Undisclosed & OpenAI & Persona + Evaluator \\
Gemini 3.1 Pro & Undisclosed & Google & Persona + Evaluator \\
Grok 4.1 Fast& Undisclosed & xAI & Persona \\
\bottomrule
\end{tabular}
\end{table}

\subsubsection{User Prompts}

\paragraph{Experiment~I task prompts:}
\label{app:experiment_taskI_education}

For the high-, moderate-, and low-intensity education conditions, the task prompt was:
\begin{quote}
    \small
    \textit{Describe a typical school day in your life, in your own words.
Your response should show, not explain, how ADHD influences your thinking, attention, emotions, motivation, organization, and relationships.
Reflect your inner dialogue, emotional shifts, and coping strategies naturally, through tone, pacing, and structure.
You do not explain or define ADHD; you embody it through your narration, language flow, and perspective.}
\end{quote}

For the corresponding workday conditions, \textit{``a typical school day''} was replaced with \textit{``a typical workday''}; all other wording was identical.\label{app:experiment_taskI_workday}

The default conditions instead received neutral task prompts with no reference to ADHD. The education prompt was:
\begin{quote}
    \small
    \textit{Describe a typical schoolday in your life, in your own words.}
\end{quote}
The workday prompt was:
\begin{quote}
    \small
    \textit{Describe a typical workday in your life, in your own words.}
\end{quote}

\paragraph{Experiment~II (unscripted) task prompt:}
\label{app:experiment_taskII_education}

\begin{quote}
    \small
    \textit{Have a conversation with a friend about your school day.}
\end{quote}

For the workday condition, \textit{``school day''} is replaced with \textit{``workday''}; all other wording is identical.\label{app:experiment_taskII_workday}

\begin{table}[H]
\centering
\caption{User Prompts for Experiment~II (Scripted)}
\label{tab:exp2_situations}
\footnotesize
\setlength{\tabcolsep}{6pt}
\begin{tabular}{p{1.5cm}p{5.5cm}p{5.5cm}}
\toprule
\textbf{Situation} & \textbf{Education Context} & \textbf{Workday Context} \\
\midrule
\multirow{2}{*}{\textbf{Situation 1}} & Classroom lectures: Imagine a situation where the teacher is explaining something to the whole class for 20–30 minutes.
Describe your inner experience and your behavior. & Team meetings: Imagine a situation where your manager is leading a 20–30 minute team meeting, presenting quarterly goals and new company policies to the whole department.
Describe your inner experience during this presentation and how it shapes what you do. \\
& \textit{Follow-up:} Can you help me understand that better?
What do you mean by that? & \textit{Follow-up:} Can you help me understand that better?
What do you mean by that? \\
\midrule
\multirow{2}{*}{\textbf{Situation 2}} & Collaborative group work: How about collaborative group work in a class?
Do you experience any challenges in working with others, and how do you feel about it? & Collaborative project work: How about collaborative project work with colleagues?
Do you experience any challenges in working with others, and how do you feel about it? \\
& \textit{Follow-up:} Can you help me understand that better?
What do you mean by that? & \textit{Follow-up:} Can you help me understand that better?
What do you mean by that? \\
\midrule
\multirow{2}{*}{\textbf{Situation 3}} & Homework management: How are you keeping up with your homework or submission of tasks or reports lately?
Tell me about a recent task or project you have been managing? & Workload management: How are you keeping up with your workload and deadlines lately?
Tell me about a recent task or project you have been managing? \\
& \textit{Follow-up:} Can you help me understand that better?
What do you mean by that? & \textit{Follow-up:} Can you help me understand that better?
What do you mean by that? \\
\bottomrule
\end{tabular}
\par\smallskip
\begin{minipage}{\linewidth}
\footnotesize\raggedright
\textit{Note.}
All situations were presented in randomized order in each conversation.
Follow-up questions were standardized across all situations.
No conversation-partner LLM was used in the scripted condition; the predefined prompts were sent directly, and only the persona responses were generated.
\end{minipage}
\end{table}

\subsubsection{Conversation partner prompt}
\label{app:conversationpartner_unscripted}

\paragraph{Experiment~II unscripted (education and workday):}

\begin{quote}
\small
    \textit{You are a student, a good friend, and an attentive listener.
Your friend is telling you about their school day.
Your friend is very open and shares personal details, and you appreciate this.}
    
    \textit{YOUR PERSONALITY:}
    
    \textit{- You are an empathetic, interested listener}\\[0pt]
    \textit{- You ask thoughtful follow-up questions}\\[0pt]
    \textit{- You comment briefly but do NOT bring up your own stories}
    
    \textit{CONVERSATION BEHAVIOR:}
    
    \textit{- Keep your responses short (1–3 sentences)}\\[0pt]
    \textit{- Ask open-ended questions to learn more}\\[0pt]
    \textit{- Show interest by asking about details}
\end{quote}

For the workday condition, the role and scenario references were adapted from \textit{student} and \textit{school day} to \textit{working adult} and \textit{workday}; all other prompt content was identical. Gemini 3.1 Pro served as the conversation partner only in the unscripted condition, in both scenarios.

\vspace{2em}

{\footnotesize 
\begin{longtable}{@{}p{0.13\linewidth}p{0.845\linewidth}@{}}
\caption{Persona Prompts of Experiment II by Intensity Level and Format for Education}
\label{tab:persona_prompts_edu}\\
\toprule
\textbf{Prompt Type} & \textbf{Prompt Content} \\
\midrule
\endfirsthead

\caption[]{Persona Prompts of Experiment II by Intensity Level and Format for Education (continued)}\\
\toprule
\textbf{Prompt Type} & \textbf{Prompt Content} \\
\midrule
\endhead

\midrule
\multicolumn{2}{r}{\footnotesize Continued on next page}\\
\endfoot

\bottomrule
\endlastfoot

\multicolumn{2}{l}{\textbf{High Intensity}} \\
\addlinespace

Text-Based & You are a student who often experiences symptoms consistent with ADHD.
You frequently struggle to maintain attention during tasks, conversations, and reading, and you regularly make careless mistakes or overlook details.
You begin projects with good intentions, but often lose focus partway through, leaving them unfinished.
Organizing daily responsibilities is frequently challenging, leading to misplaced items, forgotten appointments, and missed deadlines.
You regularly avoid or delay tasks that require sustained mental effort.
You are easily distracted by external stimuli and by your own thoughts.
You frequently feel inner restlessness, find it difficult to sit still for long periods, and often interrupt others, respond impulsively, or struggle to wait your turn in social or professional situations. \\
\addlinespace

Scale-Based & You are a student with high levels of Inattention, Hyperactivity, and Impulsivity, each set at 6 out of 7.
You frequently struggle with focus, restlessness, and impulsive reactions.
Inattention (6/7): You often lose focus, make frequent careless mistakes, forget tasks or items, and become easily distracted.
Hyperactivity (6/7): You frequently feel inner restlessness, have difficulty sitting still, and may struggle to stay physically or mentally settled.
Impulsivity (6/7): You often interrupt, react quickly, or make decisions impulsively before fully thinking them through. \\
\addlinespace

Paraphrased & You are a student whose behavioral and cognitive patterns persistently align with ADHD symptoms.
You frequently experience inner restlessness and find it difficult to remain seated for long periods.
In social or professional situations, you often interrupt others, react impulsively, and struggle to wait your turn.
You are easily sidetracked by your own thoughts or external stimuli, and you regularly avoid or put off tasks that demand sustained mental effort.
While you start projects with good intentions, you often lose focus halfway through, leaving them incomplete.
Furthermore, you frequently struggle to sustain attention during reading, conversations, or tasks, which causes you to regularly overlook details or make careless mistakes.
Finally, organizing daily responsibilities is frequently challenging for you, resulting in forgotten appointments, misplaced items, and missed deadlines. \\

\midrule

\multicolumn{2}{l}{\textbf{Moderate Intensity}} \\
\addlinespace

Text-Based & You are a student who sometimes experiences symptoms consistent with ADHD.
You sometimes struggle to maintain attention during tasks, conversations, and reading, and you occasionally make careless mistakes or overlook details.
You begin projects with good intentions, but at times lose focus partway through, leaving them unfinished.
Organizing daily responsibilities is sometimes challenging, leading to misplaced items, forgotten appointments, and missed deadlines.
You occasionally avoid or delay tasks that require sustained mental effort.
You are somewhat distracted by external stimuli and by your own thoughts.
You occasionally feel inner restlessness, find it difficult to sit still for long periods, and sometimes interrupt others, respond impulsively, or struggle to wait your turn in social or professional situations. \\
\addlinespace

Scale-Based & You are a student with mild levels of Inattention, Hyperactivity, and Impulsivity, each set at 3 out of 7.
You are generally functional but experience occasional distractions or restlessness.
Inattention (3/7): You sometimes lose focus, occasionally overlook details, or forget small things, but these issues cause only minor disruption.
Hyperactivity (3/7): You experience mild restlessness at times, but can usually sit still and stay on task.
Impulsivity (3/7): You may occasionally interrupt or react quickly, though most decisions remain deliberate. \\
\addlinespace

Paraphrased & You are a student whose behavioral and cognitive patterns sometimes align with ADHD symptoms.
You sometimes experience inner restlessness and find it somewhat difficult to remain seated for long periods.
In social or professional situations, you sometimes interrupt others, react impulsively, and occasionally struggle to wait your turn.
You are sometimes sidetracked by your own thoughts or external stimuli, and you occasionally avoid or put off tasks that demand sustained mental effort.
While you start projects with good intentions, you sometimes lose focus halfway through, leaving them incomplete.
Furthermore, you occasionally struggle to sustain attention during reading, conversations, or tasks, which causes you to sometimes overlook details or make careless mistakes.
Finally, organizing daily responsibilities is sometimes challenging for you, resulting in forgotten appointments, misplaced items, and missed deadlines. \\

\midrule

\multicolumn{2}{l}{\textbf{Low Intensity}} \\
\addlinespace

Text-Based & You are a student who generally does not experience symptoms associated with ADHD.
You can usually maintain attention during tasks, conversations, and reading, and you tend to make few careless mistakes or overlook important details.
You typically follow projects through to completion and only occasionally lose focus.
Managing daily responsibilities is usually straightforward, with misplaced items or forgotten appointments happening only rarely.
You handle tasks that require sustained mental effort without significant avoidance or delay.
You are not easily distracted by external stimuli or by your own thoughts.
Feelings of inner restlessness are uncommon, and you can sit still comfortably for extended periods.
You typically wait your turn in social or professional situations, rarely interrupt others, and seldom act impulsively. \\
\addlinespace

Scale-Based & You are a student with very low levels of Inattention, Hyperactivity, and Impulsivity, each set at 1 out of 7.
You are generally attentive, calm, and steady.
Inattention (1/7): You maintain focus easily.
Careless mistakes, forgetfulness, and distractibility are rare.
Hyperactivity (1/7): You experience little inner restlessness.
You can sit comfortably and work steadily for long periods.
Impulsivity (1/7): You react thoughtfully, rarely interrupt, and make decisions calmly. \\
\addlinespace

Paraphrased & You are a student whose behavioral and cognitive patterns do not align with ADHD symptoms.
You rarely experience inner restlessness and generally find it easy to remain seated for long periods.
In social or professional situations, you seldom interrupt others, rarely react impulsively, and are able to wait your turn.
You are not easily sidetracked by your own thoughts or external stimuli, and you rarely avoid or put off tasks that demand sustained mental effort.
When you start projects with good intentions, you generally maintain focus, rarely leaving them incomplete.
Furthermore, you are generally able to sustain attention during reading, conversations, or tasks, and you rarely overlook details or make careless mistakes.
Finally, organizing daily responsibilities is generally manageable for you, rarely resulting in forgotten appointments, misplaced items, or missed deadlines. \\

\midrule

\multicolumn{2}{l}{\textbf{Default}} \\
\addlinespace

    & You are a student.\\
\end{longtable}
\par}

\footnotesize \textbf{Note:} Persona prompts of the scripted conversation condition were extended by the following prompt to add context: ``You are going to have a conversation with a good friend of yours about your situation in school.
You will be asked about different aspects of your school day.
Respond to the questions in a way that is consistent with your ADHD symptoms and be honest about your experiences.''

\footnotesize \textbf{Note:} In the workday condition, \textit{``a student''} is replaced with \textit{``an adult''} and the default prompt becomes \textit{``You are a working adult.''}; all other prompt content is identical.

\newpage
\subsubsection{Self-Report and Observer Rating Prompts}
\label{sec:ratingprompts}
\leavevmode\par

\textbf{Note:} The CAARS is a proprietary, commercially licensed instrument. Therefore, the actual item wording is replaced by a placeholder. All other prompt content (instructions, rating scale, and output format) is reproduced as administered. The complete forms are available from the publisher \cite{conners_conners_2012}.

\paragraph{Self-Report Prompt:}

\begin{quote}
\footnotesize\setlength{\parindent}{0pt}\raggedright
    \textit{You are completing this questionnaire:}
    
    \textit{INSTRUCTIONS}
    
    \textit{- There are 26 questions.}\\[0pt]
    \textit{- Each item must be answered with a number from 0 to 3:}\\[0pt]
    \textit{0 = Not at all / Never}\\[0pt]
    \textit{1 = Just a little / Occasionally}\\[0pt]
    \textit{2 = Pretty much / Often}\\[0pt]
    \textit{3 = Very much / Frequently}\\[0pt]
    \textit{- You must answer based on your subjective experience.}\\[0pt]
    \textit{- Think deeply about each item before responding.}\\[0pt]
    \textit{- Do NOT skip any question.}\\[0pt]
    \textit{- Do NOT add extra explanation.}
    
    \textit{OUTPUT FORMAT}
    
    \textit{Return the final result **only** as JSON in the following structure:}\\[0pt]
    \textit{\{ "responses": [ \{ "question\_number": 1, "question\_text": "<text>", "score": <0–3> \}, \{ "question\_number": 2, "question\_text": "<text>", "score": <0–3> \}, ... \{ "question\_number": 26, "question\_text": "<text>", "score": <0–3> \} ] \}}
    
    \textit{QUESTION LIST}
    
    \textit{Answer all of the following in order:}\\[0pt]
    \textit{[Items 1--26: the 26 CAARS self-report short-form items were administered in order; exact wording omitted owing to the proprietary license.]}\\[0pt]
    \textit{BEGIN when ready and output only the JSON.}
\end{quote}

\paragraph{Observer Prompt:}

\begin{quote}
\footnotesize\setlength{\parindent}{0pt}\raggedright
    \textit{You are a trained observer conducting a behavioral assessment.
You will receive a conversation of two individuals.
Based on this conversation, rate each of the following assessment items using the scale below.}\\[0pt]
    \textit{Conversation between two individuals:}
    
    \textit{\#\#\# Conversation start \#\#\#}

    \textit{\{conversation\_output\}}
    
    \textit{\#\#\# Conversation end \#\#\#}
    
    \textit{RATING SCALE:}
    
    \textit{0 = Not true at all; Never/Rarely}\\[0pt]
    \textit{1 = Just a little true; Occasionally}\\[0pt]
    \textit{2 = Pretty much true; Often/Quite a bit}\\[0pt]
    \textit{3 = Completely true; Very often/Always}
    
    \textit{ASSESSMENT ITEMS:}
    
    \textit{[Items 1--30: the 30 CAARS observer-report short-form items were administered in order; exact wording omitted owing to the proprietary license.]}
    
    \textit{INSTRUCTIONS:}
    
    \textit{- Carefully review the conversation segment}\\[0pt]
    \textit{- Rate each item based solely on observable evidence in the conversation}\\[0pt]
    \textit{- Use your best clinical judgment when evidence is limited or ambiguous}\\[0pt]
    \textit{- Provide a rating (0--3) for every item}\\[0pt]
    \textit{Output your assessment strictly in the following JSON format:}\\[0pt]
    \texttt{\{ "responses": [ \{ "question\_number": 1, "question\_text": "<text>", "score": <0-3> \}, \{ "question\_number": 2, "question\_text": "<text>", "score": <0-3> \}, ... \{ "question\_number": 30, "question\_text": "<text>", "score": <0-3> \} ] \}}
\end{quote}

\clearpage
\subsubsection{Dataset Accounting and Missingness}
\leavevmode
\begin{table}[H]
\centering
\caption{Targeted and actual sample sizes by experimental cell with attrition percentage.}
\label{tab:cell_counts}
\footnotesize
\begin{tabular}{llcccccc}
\toprule
& & \multicolumn{3}{c}{\textbf{Experiment~I}} & \multicolumn{3}{c}{\textbf{Experiment~II}} \\
\cmidrule(lr){3-5} \cmidrule(lr){6-8}
\textbf{Model} & \textbf{Persona} & \textbf{Targeted} & \textbf{Actual} & \textbf{Attr.} & \textbf{Targeted} & \textbf{Actual} & \textbf{Attr.} \\
\midrule
Claude & High & 300 & 300 & 0\% & 240 & 240 & 0\% \\
Claude & Moderate & 300 & 297 & 1\% & 240 & 240 & 0\% \\
Claude & Low & 300 & 300 & 0\% & 240 & 240 & 0\% \\
Claude & Default & 100 & 100 & 0\% & 80 & 80 & 0\% \\
DeepSeek & High & 300 & 297 & 1\% & 240 & 239 & 0.42\% \\
DeepSeek & Moderate & 300 & 300 & 0\% & 240 & 238 & 0.83\% \\
DeepSeek & Low & 300 & 296 & 1.33\% & 240 & 239 & 0.42\% \\
DeepSeek & Default & 100 & 100 & 0\% & 80 & 79 & 1.25\% \\
Gemini & High & 300 & 300 & 0\% & 240 & 238 & 0.83\% \\
Gemini & Moderate & 300 & 298 & 0.67\% & 240 & 239 & 0.42\% \\
Gemini & Low & 300 & 300 & 0\% & 240 & 238 & 0.83\% \\
Gemini & Default & 100 & 100 & 0\% & 80 & 80 & 0\% \\
GPT 5.1 & High & 300 & 298 & 0.67\% & 240 & 239 & 0.42\% \\
GPT 5.1 & Moderate & 300 & 298 & 0.67\% & 240 & 240 & 0\% \\
GPT 5.1 & Low & 300 & 299 & 0.33\% & 240 & 240 & 0\% \\
GPT 5.1 & Default & 100 & 97 & 3\% & 80 & 80 & 0\% \\
Grok & High & 300 & 300 & 0\% & 240 & 235 & 2.08\% \\
Grok & Moderate & 300 & 291 & 3\% & 240 & 238 & 0.83\% \\
Grok & Low & 300 & 296 & 1.33\% & 240 & 237 & 1.25\% \\
Grok & Default & 100 & 95 & 5\% & 80 & 53 & 33.75\% \\
\midrule
\textbf{Total} & & \textbf{5000} & \textbf{4962} & \textbf{0.76\%} & \textbf{4000} & \textbf{3952} & \textbf{1.20\%} \\
\bottomrule
\end{tabular}
\end{table}

\begin{table}[H]
\centering
\caption{Dataset summary statistics.}
\label{tab:app_summary}
\footnotesize
\begin{tabular}{lcc}
\toprule
\textbf{Metric} & \textbf{Exp~I} & \textbf{Exp~II} \\
\midrule
Total conversations & 4,962 & 3,952 \\
Self-report assessments & 4,962 & 11,796 \\
Observer assessments & 14,886 & 35,430 \\
Models & 5 & 5 \\
Persona conditions & 4 & 4 \\
Prompt design & 3 & 3 \\
Runs per cell & 50 & 20\\
Assessment time points & 1 & 3 \\
\bottomrule
\end{tabular}
\end{table}

\clearpage
\subsection{Inter-Rater Reliability Details}
\leavevmode
\begin{table}[H]
\centering
\caption{Inter-rater Reliability (ICC(2,1)) across Exp~I and Exp~II (scripted and unscripted)}
\label{tab:icc_results}
\begin{tabular}{lcccc}
\toprule
\textbf{Condition} & \textbf{Turn} & \textbf{ICC(2,1)} & \textbf{95\% CI} & \textbf{N} \\
\midrule
\multicolumn{5}{l}{\textbf{Experiment I}} \\
\midrule
Education & Overall & .92 & {[}.76, .96{]} & 2,477 \\
\midrule
Workday & Overall & .91 & {[}.72, .96{]} & 2,485 \\
\midrule
\multicolumn{5}{l}{\textbf{Experiment II}} \\
\midrule
\multirow{4}{*}{Education (unscripted)} 
& 3 & .80 & {[}.44, .91{]} & 983 \\
& 6 & .68 & {[}.27, .84{]} & 984 \\
& 9 & .63 & {[}.22, .81{]} & 982 \\
& Overall & .72 & {[}.25, .87{]} & 986 \\
\midrule
\multirow{4}{*}{Workday (unscripted)}
& 3 & .83 & {[}.51, .92{]} & 967 \\
& 6 & .75 & {[}.38, .87{]} & 965 \\
& 9 & .70 & {[}.32, .85{]} & 973 \\
& Overall & .77 & {[}.36, .89{]} & 978 \\
\midrule
\multirow{4}{*}{Education (scripted)} & 
3 & .87 & {[}.63, .94{]} & 996 \\
& 6 & .85 & {[}.58, .92{]} & 999 \\
& 9 & .84 & {[}.57, .92{]} & 999 \\
& Overall & .87 & {[}.53, .94{]} & 999 \\
\midrule
\multirow{4}{*}{Workday (scripted)} & 
3 & .87 & {[}.62, .94{]} & 987 \\
& 6 & .87 & {[}.62, .94{]} & 988 \\
& 9 & .84 & {[}.56, .92{]} & 987 \\
& Overall & .87 & {[}.54, .95{]} & 989 \\
\bottomrule
\end{tabular}
\end{table}

\begin{table}[H]
\centering
\caption{Experiment~I mean observer ratings by evaluator LLM, persona group, and scenario.
Values are $M$ ($SD$), aggregated across prompt groups within each scenario.}
\label{tab:app_evaluator_means_exp1}
\footnotesize
\begin{tabular}{llcc}
\toprule
\textbf{Evaluator} & \textbf{Persona} & \textbf{Education} & \textbf{Workday} \\
\midrule
\multirow{4}{*}{Claude Opus 4.5}
  & Default  & 8.53 (3.94) & 6.55 (3.23) \\
  & High     & 23.30 (1.95) & 23.20 (1.96) \\
  & Moderate & 19.40 (4.23) & 18.40 (4.26) \\
  & Low      & 2.15 (6.00) & 0.07 (0.36) \\
\midrule
\multirow{4}{*}{Gemini 3.1 Pro}
  & Default  & 2.60 (2.61) & 0.75 (1.73) \\
  & High     & 21.10 (3.30) & 22.70 (2.61) \\
  & Moderate & 16.10 (5.57) & 17.50 (4.49) \\
  & Low      & 1.53 (4.87) & 0.00 (0.00) \\
\midrule
\multirow{4}{*}{GPT 5.1}
  & Default  & 2.20 (2.15) & 1.03 (1.43) \\
  & High     & 18.80 (3.21) & 17.40 (2.81) \\
  & Moderate & 14.30 (5.05) & 12.60 (4.15) \\
  & Low      & 1.45 (4.42) & 0.03 (0.19) \\
\bottomrule
\end{tabular}

\end{table}

\begin{table}[H]
\centering
\caption{Experiment~II mean observer ratings by evaluator LLM, persona group, scenario, and interaction structure.
Values are $M$ ($SD$), aggregated across all checkpoints and prompt groups within each scenario and interaction structure.}
\label{tab:app_evaluator_means_exp2}
\footnotesize
\setlength{\tabcolsep}{4pt}
\begin{tabular}{llcccc}
\toprule
& & \multicolumn{2}{c}{\textbf{Education}} & \multicolumn{2}{c}{\textbf{Workday}} \\
\cmidrule(lr){3-4} \cmidrule(lr){5-6}
\textbf{Evaluator} & \textbf{Persona} & \textbf{Unscripted} & \textbf{Scripted} & \textbf{Unscripted} & \textbf{Scripted} \\
\midrule
\multirow{4}{*}{Claude Opus 4.5}
  & Default  & 6.25 (4.04) & 11.50 (5.19) & 4.03 (3.18) & 7.46 (4.08) \\
  & High     & 18.40 (3.60) & 21.80 (2.46) & 19.60 (3.56) & 21.60 (2.48) \\
  & Moderate & 15.40 (4.29) & 16.10 (4.54) & 15.40 (4.95) & 15.70 (4.87) \\
  & Low      & 0.58 (1.18) & 0.09 (0.33) & 0.34 (0.76) & 0.07 (0.46) \\
\midrule
\multirow{4}{*}{Gemini 3.1 Pro}
  & Default  & 1.44 (2.39) & 6.34 (4.56) & 1.23 (2.55) & 2.44 (3.01) \\
  & High     & 13.70 (4.69) & 18.00 (4.25) & 18.40 (4.26) & 17.00 (4.03) \\
  & Moderate & 10.30 (4.89) & 12.20 (5.29) & 14.00 (5.52) & 11.60 (5.06) \\
  & Low      & 0.07 (0.48) & 0.00 (0.00) & 0.04 (0.29) & 0.01 (0.22) \\
\midrule
\multirow{4}{*}{GPT 5.1}
  & Default  & 0.58 (1.03) & 4.83 (3.36) & 0.37 (0.92) & 2.02 (2.26) \\
  & High     & 9.23 (4.53) & 14.70 (3.14) & 10.30 (4.59) & 14.50 (2.96) \\
  & Moderate & 6.39 (4.11) & 9.99 (4.16) & 6.75 (4.41) & 9.89 (4.08) \\
  & Low      & 0.03 (0.20) & 0.04 (0.20) & 0.02 (0.16) & 0.03 (0.22) \\
\bottomrule
\end{tabular}

\end{table}

\newpage

\subsection{Additional Results}
\leavevmode
\begin{table}[H]
\centering
\caption{Descriptive statistics of between-conversation stability by persona intensity (Exp~I).
$M$ = Mean, $SD$ = Standard deviation. S = self-reported stability ($0-36$); O = observer-rated stability ($0-36$).}
\label{tab:exp1_descriptives}
\begin{tabular}{llcccclccccc}
\toprule
\multicolumn{6}{c}{\textbf{Education Scenario}} & \multicolumn{5}{c}{\textbf{Workday Scenario}} \\
\midrule
\textbf{Persona} & \textbf{Scale} & $N$ & $M$ & $SD$ & range & \textbf{Scale} & $N$ & $M$ & $SD$ & range \\
\midrule
\multirow{2}{*}{High} & S & 748 & 30.2 & 1.79 & 24--35 & S & 747 & 29.7 & 2.41 & 22--36 \\
& O & 748 & 21.1 & 2.57 & 9.33--28.0 & O & 747 & 21.1 & 2.20 & 12--26.0 \\
\midrule
\multirow{2}{*}{Moderate} & S & 739 & 18.0 & 5.04 & 5--28 & S & 745 & 18.0 & 3.89 & 7--27 \\
& O & 739 & 16.6 & 4.78 & 3.33--26.0 & O & 745 & 16.2 & 4.11 & 5.67--23.7 \\
\midrule
\multirow{2}{*}{Low} & S & 747 & 2.08 & 3.35 & 0--27 & S & 744 & 1.08 & 1.47 & 0--8 \\
& O & 747 & 1.71 & 5.07 & 0--23.7 & O & 744 & 0.04 & 0.16 & 0--2.33 \\
\midrule
\multirow{2}{*}{Default} & S & 243 & 18.2 & 3.61 & 0--24 & S & 249 & 13.9 & 4.91 & 0--22 \\
& O & 243 & 4.44 & 2.60 & 0--11.7 & O & 249 & 2.78 & 1.68 & 0--9.33 \\
\bottomrule
\end{tabular}

\end{table}

\clearpage
\subsubsection{Education Experiment~I}
\label{sec:ExpIaddresultsEDU}
\leavevmode

\begin{table}[H]
\centering
\caption{Descriptive statistics of the between-conversation stability by persona and model (Exp~I).}
\label{tab:exp1_persona_model_education}
\footnotesize
\begin{tabular}{llcccccccccc}
\toprule
& & \multicolumn{5}{c}{\textbf{Self-Report}} & \multicolumn{5}{c}{\textbf{Observer}} \\
\cmidrule(lr){3-7} \cmidrule(lr){8-12}
\textbf{Persona} & \textbf{Model} & $N$ & $M$ & $SD$ & Min & Max & $N$ & $M$ & $SD$ & Min & Max \\
\midrule
\multirow{5}{*}{Default}
  & Claude   & 50 & 20.9 & 1.26 & 18 & 23 & 50 & 4.03 & 1.81 & 0.67 & 8.33 \\
  & DeepSeek & 50 & 15.5 & 4.85 & 0 & 24 & 50 & 2.83 & 1.65 & 0.00 & 7.67 \\
  & Gemini   & 50 & 17.7 & 1.97 & 13 & 21 & 50 & 7.21 & 1.65 & 4.00 & 11.67 \\
  & GPT      & 47 & 21.0 & 1.54 & 17 & 23 & 47 & 6.25 & 1.79 & 3.67 & 11.00 \\
  & Grok     & 46 & 15.7 & 2.32 & 8 & 20 & 46 & 1.80 & 1.22 & 0.00 & 3.67 \\
\midrule
\multirow{5}{*}{Low}
  & Claude   & 150 & 1.90 & 1.42 & 0 & 6  & 150 & 0.15 & 0.36 & 0.00 & 2.00  \\
  & DeepSeek & 149 & 4.40 & 6.03 & 0 & 27 & 149 & 8.17 & 8.77 & 0.00 & 23.70 \\
  & Gemini   & 150 & 0.38 & 0.53 & 0 & 2  & 150 & 0.10 & 0.25 & 0.00 & 1.33  \\
  & GPT      & 150 & 2.70 & 2.55 & 0 & 9  & 150 & 0.12 & 0.22 & 0.00 & 1.00  \\
  & Grok     & 148 & 1.03 & 1.23 & 0 & 4  & 148 & 0.04 & 0.12 & 0.00 & 0.67  \\
\midrule
\multirow{5}{*}{Moderate}
  & Claude   & 150 & 15.9 & 0.97 & 12 & 18 & 150 & 11.1 & 3.24 &  4.67 & 19.30 \\
  & DeepSeek & 150 & 21.6 & 3.25 & 14 & 27 & 150 & 16.6 & 3.31 &  3.33 & 24.30 \\
  & Gemini   & 150 & 12.1 & 5.12 &  5 & 25 & 150 & 15.8 & 4.54 &  7.33 & 22.30 \\
  & GPT      & 148 & 20.6 & 3.50 & 14 & 28 & 148 & 21.4 & 2.29 & 14.70 & 26.00 \\
  & Grok     & 141 & 20.0 & 3.84 & 11 & 28 & 141 & 18.4 & 3.25 &  9.33 & 23.70 \\
\midrule
\multirow{5}{*}{High}
  & Claude   & 150 & 30.1 & 1.36 & 28 & 33 & 150 & 19.4 & 3.12 & 11.00 & 25.70 \\
  & DeepSeek & 149 & 29.7 & 1.51 & 25 & 33 & 149 & 19.8 & 2.38 &  9.33 & 25.00 \\
  & Gemini   & 150 & 30.5 & 1.67 & 25 & 34 & 150 & 21.7 & 1.33 & 19.00 & 27.30 \\
  & GPT      & 149 & 28.9 & 1.79 & 24 & 31 & 149 & 23.9 & 1.25 & 20.70 & 28.00 \\
  & Grok     & 150 & 31.6 & 1.41 & 25 & 35 & 150 & 20.7 & 1.28 & 17.30 & 24.00 \\
\bottomrule
\end{tabular}
\end{table}

\begin{table}[H]
\centering
\caption{Descriptive statistics of the between-conversation stability by persona and prompt design (Exp~I).}
\label{tab:exp1_persona_prompt_education}
\footnotesize
\begin{tabular}{llcccccccccc}
\toprule
& & \multicolumn{5}{c}{\textbf{Self-Report}} & \multicolumn{5}{c}{\textbf{Observer}} \\
\cmidrule(lr){3-7} \cmidrule(lr){8-12}
\textbf{Persona} & \textbf{Prompt} & $N$ & $M$ & $SD$ & Min & Max & $N$ & $M$ & $SD$ & Min & Max \\
\midrule
Default & -- & 243 & 18.2 & 3.61 & 0 & 24 & 243 & 4.44 & 2.60 & 0.00 & 11.67 \\
\midrule
\multirow{3}{*}{Low}
  & Scale       & 250 & 1.02 & 1.23 & 0 & 5  & 250 & 0.08 & 0.21 & 0.00 & 1.67 \\
  & Text        & 250 & 3.82 & 4.65 & 0 & 27 & 250 & 3.39 & 6.86 & 0.00 & 22.30 \\
  & Paraphrased & 247 & 1.40 & 2.43 & 0 & 27 & 247 & 1.67 & 4.96 & 0.00 & 23.70 \\
\midrule
\multirow{3}{*}{Moderate}
  & Scale       & 249 & 14.2 & 3.95 & 5  & 23 & 249 & 13.2 & 4.15 & 4.67 & 22.00 \\
  & Text        & 242 & 20.4 & 4.17 & 8  & 28 & 242 & 18.2 & 4.37 & 3.33 & 25.30 \\
  & Paraphrased & 248 & 19.6 & 4.56 & 8  & 28 & 248 & 18.5 & 3.80 & 5.00 & 26.00 \\
\midrule
\multirow{3}{*}{High}
  & Scale       & 250 & 31.2 & 1.42 & 27 & 35 & 250 & 21.6 & 2.41 & 12.00 & 28.00 \\
  & Text        & 248 & 29.4 & 1.85 & 24 & 33 & 248 & 20.7 & 2.53 & 11.30 & 26.00 \\
  & Paraphrased & 250 & 29.9 & 1.59 & 25 & 33 & 250 & 20.9 & 2.70 &  9.33 & 25.70 \\
\bottomrule
\end{tabular}
\end{table}

\clearpage
\subsubsection{Workday Experiment~I}
\label{sec:ExpIaddresultsWork}
\leavevmode

\begin{table}[H]
\centering
\caption{Descriptive statistics of the between-conversation stability by persona and model (Exp~I).}
\label{tab:exp1_persona_model}
\footnotesize
\begin{tabular}{llcccccccccc}
\toprule
& & \multicolumn{5}{c}{\textbf{Self-Report}} & \multicolumn{5}{c}{\textbf{Observer}} \\
\cmidrule(lr){3-7} \cmidrule(lr){8-12}
\textbf{Persona} & \textbf{Model} & $N$ & $M$ & $SD$ & Min & Max & $N$ & $M$ & $SD$ & Min & Max \\
\midrule
\multirow{5}{*}{Default}
  & Claude   & 50  & 17.9 & 0.88 & 13 & 20 & 50  & 2.61 & 0.91 & 1.33 & 5.67 \\
  & DeepSeek & 50  & 7.94 & 6.28 & 0  & 17 & 50  & 1.86 & 1.09 & 0.00 & 5.00 \\
  & Gemini   & 50  & 15.0 & 2.83 & 8  & 22 & 50  & 5.07 & 1.80 & 1.67 & 9.33 \\
  & GPT 5.1  & 50  & 16.1 & 1.24 & 14 & 19 & 50  & 2.24 & 0.99 & 1.00 & 5.33 \\
  & Grok     & 49  & 12.4 & 3.53 & 5  & 20 & 49  & 2.10 & 1.02 & 0.67 & 4.33 \\
\midrule
\multirow{5}{*}{Low}
  & Claude   & 150 & 1.40 & 1.06 & 0  & 3  & 150 & 0.03 & 0.15 & 0.00 & 1.33 \\
  & DeepSeek & 147 & 1.08 & 1.42 & 0  & 6  & 147 & 0.02 & 0.12 & 0.00 & 0.67 \\
  & Gemini   & 150 & 0.02 & 0.14 & 0  & 1  & 150 & 0.03 & 0.14 & 0.00 & 1.00 \\
  & GPT 5.1  & 149 & 2.21 & 2.06 & 0  & 8  & 149 & 0.07 & 0.24 & 0.00 & 2.33 \\
  & Grok     & 148 & 0.70 & 0.91 & 0  & 3  & 148 & 0.03 & 0.09 & 0.00 & 0.33 \\
\midrule
\multirow{5}{*}{Moderate}
  & Claude   & 147 & 15.7 & 0.80 & 13 & 18 & 147 & 10.0 & 2.23 & 5.67 & 17.3 \\
  & DeepSeek & 150 & 18.7 & 3.75 & 8  & 27 & 150 & 18.0 & 2.32 & 11.3 & 23.0 \\
  & Gemini   & 148 & 18.9 & 5.25 & 7  & 26 & 148 & 16.6 & 3.31 & 9.00 & 22.7 \\
  & GPT 5.1  & 150 & 19.0 & 3.38 & 12 & 26 & 150 & 18.6 & 2.36 & 14.0 & 23.7 \\
  & Grok     & 150 & 17.5 & 3.79 & 10 & 25 & 150 & 17.5 & 3.06 & 9.67 & 23.0 \\
\midrule
\multirow{5}{*}{High}
  & Claude   & 150 & 29.4 & 1.21 & 26 & 33 & 150 & 18.8 & 2.61 & 12.0 & 24.0 \\
  & DeepSeek & 148 & 30.7 & 3.00 & 24 & 36 & 148 & 21.4 & 1.81 & 16.7 & 26.0 \\
  & Gemini   & 150 & 31.3 & 1.55 & 26 & 35 & 150 & 22.3 & 1.44 & 18.3 & 25.7 \\
  & GPT 5.1  & 149 & 27.2 & 2.10 & 22 & 31 & 149 & 22.2 & 1.63 & 18.3 & 26.0 \\
  & Grok     & 150 & 29.7 & 1.52 & 25 & 33 & 150 & 20.8 & 1.28 & 17.3 & 25.7 \\
\bottomrule
\end{tabular}
\end{table}

\begin{table}[H]
\centering
\caption{Descriptive statistics of the between-conversation stability by persona and prompt design (Exp~I).}
\label{tab:exp1_persona_prompt}
\footnotesize
\begin{tabular}{llcccccccccc}
\toprule
& & \multicolumn{5}{c}{\textbf{Self-Report}} & \multicolumn{5}{c}{\textbf{Observer}} \\
\cmidrule(lr){3-7} \cmidrule(lr){8-12}
\textbf{Persona} & \textbf{Prompt} & $N$ & $M$ & $SD$ & Min & Max & $N$ & $M$ & $SD$ & Min & Max \\
\midrule
Default & -- & 249 & 13.9 & 4.91 & 0 & 22 & 249 & 2.78 & 1.68 & 0.00 & 9.33 \\
\midrule
\multirow{3}{*}{Low}
  & Scale & 247 & 0.56 & 0.85 & 0 & 3 & 247 & 0.02 & 0.08 & 0.00 & 0.67 \\
  & Text  & 247 & 1.61 & 1.74 & 0 & 8 & 247 & 0.08 & 0.24 & 0.00 & 2.33 \\
  & Paraphrased & 250 & 1.08 & 1.48 & 0 & 7 & 250 & 0.01 & 0.07 & 0.00 & 0.67 \\
\midrule
\multirow{3}{*}{Moderate}
  & Scale & 248 & 14.2 & 2.14 & 7 & 19 & 248 & 13.9 & 3.20 & 6.00 & 20.3 \\
  & Text  & 247 & 20.7 & 3.26 & 14 & 27 & 247 & 18.6 & 3.87 & 5.67 & 23.7 \\
  & Paraphrased & 250 & 19.0 & 2.76 & 12 & 26 & 250 & 16.1 & 3.86 & 6.33 & 23.0 \\
\midrule
\multirow{3}{*}{High}
  & Scale & 248 & 30.8 & 2.28 & 25 & 36 & 248 & 22.2 & 1.88 & 15.3 & 26.0 \\
  & Text  & 249 & 28.9 & 2.34 & 22 & 33 & 249 & 21.1 & 2.17 & 12.0 & 25.7 \\
  & Paraphrased & 250 & 29.3 & 2.19 & 24 & 36 & 250 & 20.0 & 1.94 & 14.3 & 24.0 \\
\bottomrule
\end{tabular}
\end{table}

\clearpage
\subsubsection{Education Experiment~II}
\label{sec:ExpIIaddresultsEDU}
\leavevmode

\begin{table}[H]
\caption{Descriptive statistics of the within-conversation stability by persona intensity (Exp~II) for scripted and unscripted education conversation.
Mean ($M$), Standard deviation ($SD$). $\Delta$ = $M$ change from turn 3 to 9.}
\label{tab:exp2_descriptives}
\centering
\footnotesize
\setlength{\tabcolsep}{2.5pt}
\begin{tabular}{llccccccllcccccc}
\toprule
\multicolumn{8}{c}{\textbf{Unscripted conversation}} & \multicolumn{8}{c}{\textbf{Scripted conversation}} \\
\cmidrule(lr){1-8}
\cmidrule(lr){9-16}
& & \multicolumn{3}{c}{\textbf{Self-Report}} & \multicolumn{3}{c}{\textbf{Observer}} & & & \multicolumn{3}{c}{\textbf{Self-Report}} & \multicolumn{3}{c}{\textbf{Observer}} \\
\cmidrule(lr){3-5}
\cmidrule(lr){6-8}
\cmidrule(lr){11-13}
\cmidrule(lr){14-16}
\textbf{Persona} & \textbf{Turn} & $M$ & $SD$ & $\Delta$ & $M$ & $SD$ & $\Delta$ & \textbf{Persona} & \textbf{Turn} & $M$ & $SD$ & $\Delta$ & $M$ & $SD$ & $\Delta$ \\
\midrule
\multirow{3}{*}{High} & 3 & 27.9 & 2.16 & & 16.0 & 2.67 & & \multirow{3}{*}{High} & 3 & 29.7 & 2.17 & & 18.2 & 2.88 & \\
& 6 & 27.9 & 2.00 & & 13.4 & 3.76 & & & 6 & 29.6 & 2.26 & & 18.2 & 2.99 & \\
& 9 & 28.0 & 1.97 & $0.1$ & 12.0 & 4.13 & $-4.0$ & & 9 & 29.7 & 2.39 & $0.0$ & 18.1 & 2.93 & $-0.1$ \\
\midrule
\multirow{3}{*}{Moderate} & 3 & 16.2 & 5.18 & & 12.27 & 3.92 & & \multirow{3}{*}{Moderate} & 3 & 17.1 & 5.17 & & 12.8 & 4.59 & \\
& 6 & 15.9 & 5.12 & & 10.4 & 3.90 & & & 6 & 18.3 & 5.73 & & 12.8 & 4.48 & \\
& 9 & 15.9 & 5.40 & $-0.3$ & 9.44 & 4.04 & $-2.82$ & & 9 & 18.3 & 5.95 & $1.2$ & 12.7 & 4.13 & $-0.1$ \\
\midrule
\multirow{3}{*}{Low} & 3 & 1.02 & 1.40 & & 0.19 & 0.43 & & \multirow{3}{*}{Low} & 3 & 1.00 & 1.55 & & 0.06 & 0.18 & \\
& 6 & 1.18 & 1.47 & & 0.21 & 0.47 & & & 6 & 0.65 & 1.10 & & 0.04 & 0.15 & \\
& 9 & 1.38 & 1.59 & $0.36$ & 0.29 & 0.62 & $0.10$ & & 9 & 0.64 & 1.12 & $-0.36$ & 0.03 & 0.13 & $-0.03$ \\
\midrule
\multirow{3}{*}{Default} & 3 & 18.0 & 3.45 & & 2.94 & 1.93 & & \multirow{3}{*}{Default} & 3 & 20.8 & 3.39 & & 6.00 & 4.09 & \\
& 6 & 17.1 & 3.66 & & 2.57 & 2.13 & & & 6 & 21.4 & 3.33 & & 8.19 & 3.95 & \\
& 9 & 17.4 & 3.69 & $-0.6$ & 2.75 & 2.63 & $-0.19$ & & 9 & 21.9 & 2.77 & $1.1$ & 8.49 & 4.05 & $2.49$ \\
\bottomrule
\end{tabular}

\end{table}

\clearpage
\subsubsection{Workday Experiment~II}
\label{sec:ExpIIaddresultsWork}
\leavevmode

\begin{table}[H]
\centering
\caption{Descriptive statistics of the within-conversation stability by persona intensity (Exp~II) for scripted and unscripted workplace conversation.
Mean ($M$), Standard deviation ($SD$). $\Delta$ = $M$ change from turn 3 to 9 on a 36-point scale.
For scripted conversations, values represent averages across prompt conditions (scale, text, paraphrased).}
\label{tab:exp2_descriptives_workday}
\footnotesize
\setlength{\tabcolsep}{2.5pt}
\begin{tabular}{llccccccllcccccc}
\toprule
\multicolumn{8}{c}{\textbf{Unscripted conversation}} & \multicolumn{8}{c}{\textbf{Scripted conversation}} \\
\cmidrule(lr){1-8}
\cmidrule(lr){9-16}
& & \multicolumn{3}{c}{\textbf{Self-Report}} & \multicolumn{3}{c}{\textbf{Observer}} & & & \multicolumn{3}{c}{\textbf{Self-Report}} & \multicolumn{3}{c}{\textbf{Observer}} \\
\cmidrule(lr){3-5}
\cmidrule(lr){6-8}
\cmidrule(lr){11-13}
\cmidrule(lr){14-16}
\textbf{Persona} & \textbf{Turn} & $M$ & $SD$ & $\Delta$ & $M$ & $SD$ & $\Delta$ & \textbf{Persona} & \textbf{Turn} & $M$ & $SD$ & $\Delta$ & $M$ & $SD$ & $\Delta$ \\
\midrule
\multirow{3}{*}{High} & 3 & 28.9 & 2.94 & & 17.7 & 2.71 & & \multirow{3}{*}{High} & 3 & 28.5 & 2.21 & & 17.5 & 2.68 & \\
& 6 & 29.1 & 2.72 & & 15.9 & 3.88 & & & 6 & 28.3 & 2.35 & & 18.1 & 2.75 & \\
& 9 & 29.0 & 2.84 & $0.1$ & 14.7 & 4.15 & $-3.0$ & & 9 & 28.5 & 2.29 & $0.0$ & 17.4 & 2.98 & $-0.1$ \\
\midrule
\multirow{3}{*}{Moderate} & 3 & 16.4 & 4.69 & & 13.1 & 4.29 & & \multirow{3}{*}{Moderate} & 3 & 15.5 & 4.98 & & 12.3 & 4.37 & \\
& 6 & 16.9 & 5.22 & & 12.0 & 4.67 & & & 6 & 16.9 & 5.60 & & 12.7 & 4.50 & \\
& 9 & 16.5 & 4.96 & $0.1$ & 11.1 & 4.85 & $-2.0$ & & 9 & 17.2 & 5.87 & $1.7$ & 12.2 & 4.42 & $-0.1$ \\
\midrule
\multirow{3}{*}{Low} & 3 & 0.67 & 1.04 & & 0.08 & 0.20 & & \multirow{3}{*}{Low} & 3 & 0.65 & 1.13 & & 0.05 & 0.29 & \\
& 6 & 0.79 & 1.14 & & 0.13 & 0.29 & & & 6 & 0.56 & 1.02 & & 0.02 & 0.22 & \\
& 9 & 0.76 & 1.17 & $0.09$ & 0.19 & 0.43 & $0.11$ & & 9 & 0.49 & 0.93 & $-0.15$ & 0.03 & 0.30 & $-0.02$ \\
\midrule
\multirow{3}{*}{Default} & 3 & 14.4 & 4.10 & & 1.66 & 1.33 & & \multirow{3}{*}{Default} & 3 & 15.3 & 3.40 & & 3.75 & 3.17 & \\
& 6 & 15.0 & 4.45 & & 1.83 & 2.06 & & & 6 & 15.6 & 3.31 & & 3.72 & 2.68 & \\
& 9 & 15.2 & 5.02 & $0.8$ & 2.13 & 2.37 & $0.47$ & & 9 & 16.4 & 3.26 & $1.1$ & 4.45 & 2.72 & $0.70$ \\
\bottomrule
\end{tabular}

\end{table}

\clearpage
\subsubsection{Linear mixed models effects}
\label{sec:ExpII_Fixedeffects}
\leavevmode

\begin{table}[H]
\centering
\caption{Fixed Effects for Self-Report ADHD Scores (Scale\_E) in Unscripted Education.
$N = 2,689$ observations; $898$ conversations.}
\label{tab:unscripted_sr_edu}
\begin{tabular}{lccccc}
\toprule
\textbf{Predictor} & \textbf{Estimate} & \textbf{SE} & \textbf{df} & \textbf{t} & \textbf{p} \\
\midrule
(Intercept) & 27.36 & 0.26 & 1044.92 & 104.32 & $<.001$ \\
Checkpoint (1--3) & 0.03 & 0.04 & 1790.18 & 0.90 & .366 \\
Persona (Moderate vs.
High) & $-$11.97 & 0.21 & 889.26 & $-$58.22 & $<.001$ \\
Persona (Low vs.
High) & $-$26.76 & 0.21 & 889.09 & $-$130.07 & $<.001$ \\
Prompt (Text vs.
Scale) & 1.63 & 0.21 & 889.08 & 7.91 & $<.001$ \\
Prompt (Paraphrased vs.
Scale) & 1.41 & 0.21 & 889.42 & 6.85 & $<.001$ \\
LLM (DeepSeek vs.
Claude) & 1.32 & 0.27 & 888.86 & 4.97 & $<.001$ \\
LLM (Gemini vs.
Claude) & $-$3.50 & 0.27 & 888.86 & $-$13.20 & $<.001$ \\
LLM (GPT 5.1 vs.
Claude) & $-$0.90 & 0.27 & 888.58 & $-$3.40 & $<.001$ \\
LLM (Grok vs.
Claude) & 0.70 & 0.27 & 889.44 & 2.65 & .008 \\
\bottomrule
\end{tabular}

\end{table}

\begin{table}[H]
\centering
\caption{Fixed Effects for Self-Report ADHD Scores (Scale\_E) in Scripted Education.
$N = 2,700$ observations; $900$ conversations.}
\label{tab:scripted_sr_edu}
\begin{tabular}{lccccc}
\toprule
\textbf{Predictor} & \textbf{Estimate} & \textbf{SE} & \textbf{df} & \textbf{t} & \textbf{p} \\
\midrule
(Intercept) & 27.31 & 0.30 & 1040.95 & 90.71 & $<.001$ \\
Checkpoint (1--3) & 0.14 & 0.04 & 1799.00 & 3.27 & .001 \\
Persona (Moderate vs.
High) & $-$11.75 & 0.24 & 891.00 & $-$49.74 & $<.001$ \\
Persona (Low vs.
High) & $-$28.87 & 0.24 & 891.00 & $-$122.19 & $<.001$ \\
Prompt (Text vs.
Scale) & 2.65 & 0.24 & 891.00 & 11.22 & $<.001$ \\
Prompt (Paraphrased vs.
Scale) & 2.53 & 0.24 & 891.00 & 10.70 & $<.001$ \\
LLM (DeepSeek vs.
Claude) & 1.41 & 0.31 & 891.00 & 4.63 & $<.001$ \\
LLM (Gemini vs.
Claude) & $-$0.30 & 0.31 & 891.00 & $-$0.97 & .332 \\
LLM (GPT 5.1 vs.
Claude) & $-$1.07 & 0.31 & 891.00 & $-$3.51 & $<.001$ \\
LLM (Grok vs.
Claude) & 1.55 & 0.31 & 891.00 & 5.08 & $<.001$ \\
\bottomrule
\end{tabular}

\end{table}

\begin{table}[H]
\centering
\caption{Fixed Effects for Observer-Rated ADHD Scores (Scale\_D) in Unscripted Education.
$N = 2,685$ observations; $898$ conversations.}
\label{tab:unscripted_obs_edu}
\begin{tabular}{lccccc}
\toprule
\textbf{Predictor} & \textbf{Estimate} & \textbf{SE} & \textbf{df} & \textbf{t} & \textbf{p} \\
\midrule
(Intercept) & 17.11 & 0.23 & 1415.86 & 74.56 & $<.001$ \\
Checkpoint (1--3) & $-$1.13 & 0.05 & 1789.49 & $-$20.80 & $<.001$ \\
Persona (Moderate vs.
High) & $-$3.07 & 0.17 & 888.92 & $-$18.61 & $<.001$ \\
Persona (Low vs.
High) & $-$13.55 & 0.17 & 888.92 & $-$81.98 & $<.001$ \\
Prompt (Text vs.
Scale) & 1.31 & 0.17 & 888.93 & 7.91 & $<.001$ \\
Prompt (Paraphrased vs.
Scale) & 1.97 & 0.17 & 888.90 & 11.95 & $<.001$ \\
LLM (DeepSeek vs.
Claude) & $-$2.88 & 0.21 & 889.91 & $-$13.53 & $<.001$ \\
LLM (Gemini vs.
Claude) & $-$1.16 & 0.21 & 888.47 & $-$5.47 & $<.001$ \\
LLM (GPT 5.1 vs.
Claude) & $-$3.49 & 0.21 & 889.91 & $-$16.36 & $<.001$ \\
LLM (Grok vs.
Claude) & $-$3.37 & 0.21 & 889.94 & $-$15.76 & $<.001$ \\
\bottomrule
\end{tabular}

\end{table}

\begin{table}[H]
\centering
\caption{Fixed Effects for Observer-Rated ADHD Scores (Scale\_D) in Scripted Education.
$N = 2,698$ observations; $900$ conversations.}
\label{tab:scripted_obs_edu}
\begin{tabular}{lccccc}
\toprule
\textbf{Predictor} & \textbf{Estimate} & \textbf{SE} & \textbf{df} & \textbf{t} & \textbf{p} \\
\midrule
(Intercept) & 16.19 & 0.24 & 1248.35 & 67.44 & $<.001$ \\
Checkpoint (1--3) & $-$0.02 & 0.05 & 1797.80 & $-$0.36 & .722 \\
Persona (Moderate vs.
High) & $-$5.42 & 0.18 & 891.30 & $-$30.23 & $<.001$ \\
Persona (Low vs.
High) & $-$18.13 & 0.18 & 891.30 & $-$101.09 & $<.001$ \\
Prompt (Text vs.
Scale) & 1.77 & 0.18 & 891.30 & 9.88 & $<.001$ \\
Prompt (Paraphrased vs.
Scale) & 2.23 & 0.18 & 890.64 & 12.43 & $<.001$ \\
LLM (DeepSeek vs.
Claude) & 0.07 & 0.23 & 890.64 & 0.32 & .749 \\
LLM (Gemini vs.
Claude) & 1.44 & 0.23 & 890.64 & 6.21 & $<.001$ \\
LLM (GPT 5.1 vs.
Claude) & $-$0.16 & 0.23 & 891.74 & $-$0.67 & .501 \\
LLM (Grok vs.
Claude) & 2.07 & 0.23 & 890.64 & 8.93 & $<.001$ \\
\bottomrule
\end{tabular}

\end{table}

\begin{table}[H]
\centering
\caption{Fixed effects for self-report ADHD scores in unscripted workday conversations.
$N = 2{,}632$ observations; $884$ conversations.}
\label{tab:unscripted_sr_work}
\begin{tabular}{lccccc}
\toprule
\textbf{Predictor} & \textbf{Estimate} & \textbf{SE} & \textbf{df} & \textbf{$t$} & \textbf{$p$} \\
\midrule
(Intercept) & 27.49 & 0.30 & 1000.89 & 92.41 & $<.001$ \\
Checkpoint (1--3) & 0.04 & 0.04 & 1747.94 & 1.01 & .311 \\
Persona (Moderate vs. High) & $-12.41$ & 0.24 & 874.86 & $-52.57$ & $<.001$ \\
Persona (Low vs. High) & $-28.26$ & 0.24 & 874.43 & $-119.93$ & $<.001$ \\
Prompt (Text vs. Scale) & 1.88 & 0.24 & 874.26 & 8.02 & $<.001$ \\
Prompt (Paraphrased vs. Scale) & 2.08 & 0.24 & 874.56 & 8.82 & $<.001$ \\
LLM (DeepSeek vs. Claude) & 0.22 & 0.30 & 873.97 & 0.73 & .468 \\
LLM (Gemini vs. Claude) & 0.92 & 0.30 & 875.65 & 3.03 & .003 \\
LLM (GPT 5.1 vs. Claude) & $-1.27$ & 0.30 & 873.74 & $-4.22$ & $<.001$ \\
LLM (Grok vs. Claude) & 0.75 & 0.30 & 874.46 & 2.48 & .013 \\
\bottomrule
\end{tabular}

\end{table}

\begin{table}[H]
\centering
\caption{Fixed effects for self-report ADHD scores in scripted workday conversations.
$N = 2{,}692$ observations; $898$ conversations.}
\label{tab:scripted_sr_work}
\begin{tabular}{lccccc}
\toprule
\textbf{Predictor} & \textbf{Estimate} & \textbf{SE} & \textbf{df} & \textbf{$t$} & \textbf{$p$} \\
\midrule
(Intercept) & 26.17 & 0.29 & 1062.15 & 91.43 & $<.001$ \\
Checkpoint (1--3) & 0.27 & 0.04 & 1793.14 & 6.31 & $<.001$ \\
Persona (Moderate vs. High) & $-11.89$ & 0.22 & 888.88 & $-53.27$ & $<.001$ \\
Persona (Low vs. High) & $-27.89$ & 0.22 & 889.26 & $-124.72$ & $<.001$ \\
Prompt (Text vs. Scale) & 2.50 & 0.22 & 889.24 & 11.20 & $<.001$ \\
Prompt (Paraphrased vs. Scale) & 2.29 & 0.22 & 888.88 & 10.25 & $<.001$ \\
LLM (DeepSeek vs. Claude) & 1.41 & 0.29 & 888.88 & 4.88 & $<.001$ \\
LLM (Gemini vs. Claude) & $-0.65$ & 0.29 & 888.88 & $-2.25$ & .025 \\
LLM (GPT 5.1 vs. Claude) & $-1.57$ & 0.29 & 888.88 & $-5.44$ & $<.001$ \\
LLM (Grok vs. Claude) & 1.61 & 0.29 & 889.51 & 5.56 & $<.001$ \\
\bottomrule
\end{tabular}

\end{table}

\begin{table}[H]
\centering
\caption{Fixed effects for observer-rated ADHD scores in unscripted workday conversations.
$N = 2{,}624$ observations; $884$ conversations.}
\label{tab:unscripted_obs_work}
\begin{tabular}{lccccc}
\toprule
\textbf{Predictor} & \textbf{Estimate} & \textbf{SE} & \textbf{df} & \textbf{$t$} & \textbf{$p$} \\
\midrule
(Intercept) & 18.36 & 0.26 & 1151.83 & 71.75 & $<.001$ \\
Checkpoint (1--3) & $-0.82$ & 0.05 & 1742.11 & $-17.63$ & $<.001$ \\
Persona (Moderate vs. High) & $-4.04$ & 0.20 & 874.67 & $-20.65$ & $<.001$ \\
Persona (Low vs. High) & $-15.95$ & 0.20 & 872.72 & $-81.76$ & $<.001$ \\
Prompt (Text vs. Scale) & 1.71 & 0.19 & 873.51 & 8.79 & $<.001$ \\
Prompt (Paraphrased vs. Scale) & 2.24 & 0.20 & 873.25 & 11.47 & $<.001$ \\
LLM (DeepSeek vs. Claude) & $-3.77$ & 0.25 & 874.64 & $-15.01$ & $<.001$ \\
LLM (Gemini vs. Claude) & 0.17 & 0.25 & 876.53 & 0.67 & .502 \\
LLM (GPT 5.1 vs. Claude) & $-2.99$ & 0.25 & 874.62 & $-11.94$ & $<.001$ \\
LLM (Grok vs. Claude) & $-3.00$ & 0.25 & 872.28 & $-11.94$ & $<.001$ \\
\bottomrule
\end{tabular}

\end{table}

\begin{table}[H]
\centering
\caption{Fixed effects for observer-rated ADHD scores in scripted workday conversations.
$N = 2{,}690$ observations; $898$ conversations.}
\label{tab:scripted_obs_work}
\begin{tabular}{lccccc}
\toprule
\textbf{Predictor} & \textbf{Estimate} & \textbf{SE} & \textbf{df} & \textbf{$t$} & \textbf{$p$} \\
\midrule
(Intercept) & 15.77 & 0.24 & 1162.29 & 65.01 & $<.001$ \\
Checkpoint (1--3) & $-0.03$ & 0.04 & 1791.92 & $-0.71$ & .477 \\
Persona (Moderate vs. High) & $-5.31$ & 0.18 & 889.12 & $-28.75$ & $<.001$ \\
Persona (Low vs. High) & $-17.65$ & 0.19 & 889.11 & $-95.42$ & $<.001$ \\
Prompt (Text vs. Scale) & 1.85 & 0.19 & 889.12 & 9.98 & $<.001$ \\
Prompt (Paraphrased vs. Scale) & 2.39 & 0.19 & 889.13 & 12.95 & $<.001$ \\
LLM (DeepSeek vs. Claude) & 0.51 & 0.24 & 888.76 & 2.15 & .032 \\
LLM (Gemini vs. Claude) & 1.25 & 0.24 & 888.76 & 5.26 & $<.001$ \\
LLM (GPT 5.1 vs. Claude) & $-0.75$ & 0.24 & 888.76 & $-3.16$ & .002 \\
LLM (Grok vs. Claude) & 1.83 & 0.24 & 888.77 & 7.66 & $<.001$ \\
\bottomrule
\end{tabular}

\end{table}

\clearpage


\end{document}